\documentclass{elsarticle}
\usepackage{amssymb,amsmath}
\usepackage{natbib}
\usepackage{url} 
\usepackage{graphicx}
\usepackage{subfigure}
\usepackage{subfloat}
\usepackage[caption=false,font=footnotesize]{subfig}
\usepackage{algorithm}
\usepackage{algpseudocode}

\algnewcommand{\LineComment}[1]{\State \(\triangleright\) #1}
\algnewcommand{\Parameter}[1]{\State #1}

\algnewcommand\algorithmicinput{\textbf{INPUT:}}
\algnewcommand\INPUT{\item[\algorithmicinput]}
\algnewcommand\algorithmicoutput{\textbf{OUTPUT:}}
\algnewcommand\OUTPUT{\item[\algorithmicoutput]}
\algnewcommand\algorithmicglobalvariables{\textbf{GLOBAL VARIABLES:}}
\algnewcommand\GLOBALVARIABLES{\item[\algorithmicglobalvariables]}
\algnewcommand\And{\textbf{and}}

\makeatletter
\newcommand{\StatexIndent}[1][3]{%
  \setlength\@tempdima{\algorithmicindent}%
  \Statex\hskip\dimexpr#1\@tempdima\relax}
\algdef{S}[WHILE]{WhileNoDo}[1]{\algorithmicwhile\ #1}%
\makeatother

\newtheorem{definition}{Definition}[section]

\begin{document}
\title{$ER$-index: a referential index for encrypted genomic databases}
\author[1]{Ferdinando Montecuollo} \ead{montecuollo@gmail.com} 
\author[2]{Giovannni Schmid\corref{cor1}} \ead{giovanni.schmid@cnr.it}
\address[1]{CRESSI, Universit\`{a} ``Luigi Vanvitelli'', Napoli, 80133 Italy}
\address[2]{ICAR, Consiglio Nazionale delle Ricerche, Napoli, 80131, Italy}

\begin{abstract}
Huge DBMSs storing genomic information are being created and engineerized 
for doing large-scale, comprehensive and in-depth analysis of human beings and 
their diseases. This paves the way for significant new approaches in medicine,
but also poses major challenges for storing, processing and transmitting such 
big amounts of data in compliance with recent regulations concerning user privacy.\\
We designed and implemented \textit{$ER$-index}, a new full-text index in minute 
space which was optimized for pattern-search on compressed and encrypted genomic 
data using a reference sequence, and that complements a previous index for 
reference-free genomics. 
Thanks to a multi-user and multiple-keys encryption model, a single $ER$-index 
can store the sequences related to a large population of individuals so that 
users may perform search operations directly on compressed data and only on the 
sequences to which they were granted access. \\
Tests performed of three different computing platforms show that the $ER$-index 
get very good compression ratios and search times, outperforming in many cases 
a reference tool that was proved nearly-optimal in time and space and does not 
implement encryption.\\ 
The $ER$-index C++ source code plus scripts and data to assess the tool performance 
are available at: \url{https://github.com/EncryptedIndexes/erindex}.
\end{abstract}


\maketitle

\section{Introduction}
Predictive, preventive, precise and participatory medicine (\textit{P4 medicine}, 
for short) are new approaches underpinned by genome sequencing that will soon be 
incorporated in our health systems. The advantages of these approaches for human 
health and wellbeing can be very significant: 
by reshaping healthcare from reactive to proactive they indeed represent the main 
answer to the progression of ``silent'' chronic diseases, which are the leading 
cause of death, disability and diminished quality of life in the developed world, 
strongly impacting the economy of many countries \cite{sagner2017p4}.

However, such new approaches pose very big computational and security challenges.

\noindent
Data management in genomics is considered a ``four-headed beast'' due to the high
computational costs concerning the lifecycle of a data set: acquisition, storage, 
distribution and analysis. According to \cite{stephens2015big}, the total amount 
of sequenced data has doubled approximately every seven months from 2005 to 2015, 
and if the growth will continue at this rate then we should approach one zettabase 
of sequence per year by 2025. Storage requirements will be much bigger: dozens 
of exabytes of storage capacity could be required by 2025 just for the human 
genomes, since for each sequenced base about 10-fold more data must be collected for 
sequencing errors, base calling and sequence alignment. Data compression tools which 
are effective for genomic data can greatly diminish these needs, but they 
do not mitigate the computational bottleneck: the original uncompressed data set 
must indeed be reconstructed before it can be analyzed.

\noindent
On the other hand, the storage and processing of human genomic data raise major 
privacy and safety concerns. 
Human genome projects were initially open access, since it was believed that there was 
no risk of identification of participants or donors, but this approach was overturned 
after \cite{homer2008resolving} realized that data from individuals could be distinguished 
in \textit{Genome Wide Association Studies} (GWAS) just using summary statistics. 
Later studies have shown that correlations among a small number of pre-specified genetic 
regions from anonymized biobanks and other public datasets can effectively identify 
individuals. For example, \cite{gymrek2013identifying} recovered surnames from personal 
genomes by profiling short tandem repeats on the Y chromosome, and then used other types 
of metadata, such as age and state, to identify the targets.\\
Despite the increasing threats posed by such kind of techniques, different regulations in
force (e.g., HIPAA \cite{HIPAA}, GDPR \cite{GDPR}) do not pose explicit privacy restrictions
on the use or disclosure of health information that has been de-identified. 
This, together with the fact that encryption is perceived as a barrier for both usability 
and performance, is arguably curbing the practice of encrypting genomic data at rest. 
As a matter of fact, storing and sharing human genome data sets with the sole protection 
of some sort of anonymization and access control has become a common practice in genomics. 
For example, the Genomic Data Commons Data Portal (https://portal.gdc.cancer.gov/) is a 
huge database managed by the NIH in which the data are compressed thanks to the GNU Gzip 
algorithm, possibly after creating a TAR archive in case of multiple files, and then stored 
in the clear and accessed thanks to authentication credentials. \\
This state of affairs exposes an increasing number of institutions and individuals to the 
disclosement and tampering of very sensitive data, with profound implications on economy
and safety. Indeed, according to some prominent data risk reports (e.g. \cite{thales2019,
identityforce2020, varonis2019}), data theft is a very frequent exploit from which no 
organization is really immune. This is because it can be a consequence of vulnerabilities 
that are difficult to face, like misconfigurations in access control policies, poor access 
credentials management and software bugs. Furthermore, data theft can be perpetrated by data 
custodians themselves for profit, especially in contexts where data are externalized (e.g., 
stored in the Cloud) 

\subsection{Encrypted and compressed full-text}
Encrypted and compressed full-text indexes are very promising tools for tackling such 
challenges and paving the way to P4 medicine.

Indexes are data structures that allow indexed data searching, thus improving pattern search performance.
Classical full-text indexes (e.g., suffix-trees and suffix-arrays) can extract any text substring 
from arbitrary data text efficiently in time; however, their use is unfeasible for large data sets, 
since they require at least 2.5 times the input text space (plus text) to achieve reasonable efficiency
\cite{navarro2001indexing}.
Introduced in a seminal work in 2000 \cite{ferragina2000opportunistic}, \textit{compressed full-text 
self-indexes} are succint data structures that carefully combine loseless compression and data indexing
techniques in order to obtain the search power of suffix arrays/trees and a space occupancy close
to the one achievable by the best known compressors. Moreover, they are named \textit{self-indexes} since  
they are able to reproduce any text portion without accessing the original text, and thus the input text 
can be discarded.
These algorithms can offer great performance improvements if compared to the approach ``decompress-and-search'' 
supported through standard compression tools for genomic data (for a recent review on such tools see 
\cite{hosseini2016survey}). 
First and foremost, they allow for searching directly on compressed data, avoiding decompress and compress 
operations each time a new search is required, which can be very costly. 
For example GDC 2 \cite{deorowicz2015gdc}, which was among the best reference-based methods considered 
in \cite{hosseini2016survey}, consumed in mean about eight minutes to compress 128 MB of the GRC reference 
assembly for human chromosome 11 and more than nine hours to compress the H.sapiens dataset (6670 GB) 
on high-end workstations \cite{hosseini2016survey, deorowicz2015gdc}. 
In addition, compressed full-text indexes operate directly in RAM searches for patterns that would 
otherwise need to recover data from the disk after decompression, thus being orders of magnitude faster.

On the other hand, the most effective countermeasure to data theft is given by modern encryption 
schemes: provided they are correctly implemented and their cryptographic parameters (especially 
private keys) are managed securely, data breaches are rendered useless by the use of encryption.
However, in the context of genomic data the choice of the cryptographic algorithms and protocols and 
their implementation have to be taken very seriously, otherwise the resulting system could be unacceptably 
inefficient, and/or not adequately protected in the long term. In this respect, it can be worthwhile to 
stress that the ``compress-then-encrypt'' approach -- in which encryption is performed with a dedicated 
algorithm after data compression -- has the drawback that one must first decrypt the index in its 
entirety before operating through it. For massive data amounts, as in case of genomic datasets, this can 
lead to big downgrades in performance; moreover, it exposes data on disk during operations, which can be 
an issue if the databases are in outsourcing or in multi-tenants environments (e.g. Cloud environments).

\subsection{Paper contribution and organization}
In the present work we introduce $ER$-index (\textit{Encrypted Referential} index), the first encrypted 
self-index based on referential Lempel-Ziv compression, and designed so that it can be the core of a 
multi-user database engine. 

\noindent
Being a self-index, the $ER$-index represents a \textit{compressive genomics} algorithm in the sense of 
\cite{loh2012compressive}: it indeed allows to search for patterns directly on compressed genomic data.
Compressive algorithms are the only tools which can cope with the aforementioned exponential grow of
genomics data. Indeed, any computational analysis running even on a constant fraction of the full genomic 
library, such as sequence search,  scales at least linearly in time with respect to the size of the library 
and therefore grows exponentially slower every year. 

\noindent
Being referential, the $ER$-index fully exploits the high similarity among human genomes, which is about
99,5\% \cite{levy2007diploid}. This allows for much better compression ratios than non-referential indices 
but without penalties of practical significance in pattern search performance. As shown by the numerical 
experiments described in a following section, for collections of 50 chromosomes the $ER$-index achieves 
compression ratios four times better and pattern search times very close to those of the reference tool, 
which is a self-index nearly optimal in time and space not implementing referential compression.

\noindent
Last but not least, the $ER$-index implements a built-in encryption algorithm which allows the decryption
of just the index blocks involved in the pattern search operation, thus saving computing time and memory
space if compared to encryption realized via an external tool. The $ER$-index \textit{multi-user encryption 
model} permits to store genomic sequences of different individuals with distinct encryption keys within the 
same index. This allows users to perform decrypt operations only on the sequences to which they were granted 
access.  

The paper is organized as follows. Section \ref{sec:relatedwork} discusses previous work
directly related to our present work. Section \ref{sec:systemandmethods} illustrates the system and service 
having at their computational core the $ER$-index, alongside with the intended application scenarios. This 
section discusses also the security advantages of our proposal versus current biobank services and possible 
alternatives approaches, on the basis of a threat model which is coherent with the application domain.
Section \ref{sec:encrypted_referential} gives an overview of the main features of $ER$-index, alongside with 
the computational methods and data structures which make possible such features. 
Sections \ref{sec:factorization}, \ref{sec:pattern_search} and \ref{sec:encryption} give details on its core 
algorithms, pointing out some important differences of our approach with respect to current computing techniques 
for genomic databases.
Section \ref{sec:results} reports and discusses the results of the tests we have run 
in order to assess the performance of our tool versus a state-of-the-art index, the 
\textit{wavelet tree $FM$-index} by \cite{grossi2003high}.
Finally, Section \ref{sec:conclusion} sums up the main features of the $ER$-index and
sketches out future work.

\section{Related work}
\label{sec:relatedwork}
The \textit{Fast index in Minute Space} ($FM$-index) introduced in \cite{ferragina2000opportunistic} 
is an example of full-text self-index whose space occupancy is a function of the entropy of the 
underlying data set: its space occupancy is decreased when the input is compressible, and this space
reduction is achieved at no significant slowdown in the query performance. The $FM$-index of a text
$T$ is composed of a binary sequence which results from the compression of the \textit{Burrows-Wheeler 
transform} (BWT) \cite{burrows1994block} of $T$, plus a set of auxiliary data structures introduced 
to speed up query performance. These data structures describe the subdivision of the binary sequence 
into data blocks of variable length, each block being related to a single fixed-length block of the BWT.
In \cite{montecuollo2017e2fm} was introduced the $E^2FM$-index, a full-text index in minute 
space which was optimized for compressing and encrypting nucleotide sequence collections, 
and for performing fast pattern-search queries on them, without the knowledge of a reference 
sequence. The $E^2FM$-index is particularly suitable for \textit{metagenomics} 
\cite{markowitz2012img} or \textit{de-novo discovery} applications, but it does not represent 
the best choice when a large collection of similar sequences is given with respect to a reference sequence, 
since it does not implement referential compression. 
Referential compression is very effective for nucleotide sequence collections related to the same 
loci for individuals of the same specie, with a fast decrease in compression ratios at increasing 
collection size. This is because a sequence can be assumed as a reference, and all other sequences 
can be expressed just encoding their differences compared to it. 
In fact, a comparison  of the results reported in Section \ref{sec:results} with those reported in 
\cite{montecuollo2017e2fm} shows that the $ER$-index allows to obtain compression ratios which are an
order of magnitude smaller than those of $E^2FM$-index for collections of fifty items. This improvement 
cames at the cost of longer times in pattern search, which are however of the order of milliseconds.

Similarly to $FM$-index, the $E^2FM$-index first processes data thanks to the Burrows-Wheeler 
transform and the Move-to-front (MTF) transform \cite{ryabko1980data, bentley1986locally}, 
after which it compresses them with the RLE0 algorithm \cite{salomon2004data}. 
The BWT approach does not seem so appropriate for referential compression as dictionary-based 
methods, thus we have adopted another compression strategy in the $ER$-index, based on LZ77 
algorithm.  

Lempel-Ziv methods are lossless, dictionary-based compression algorithms which replace 
repetitions in a string by using references of their previous occurrences. There are many 
variants, all derived from the two algorithms introduced by \cite{ziv1977universal, 
ziv1978compression} and named LZ77 and LZ78, respectively.

Most of the self-indexes inspired to the Lempel-Ziv parsing use LZ78, because the LZ78 
factorization of a text has some interesting properties which allow to design efficient 
pattern search algorithms like that of \cite{siren2008run}. 
LZ78 is faster but more complex than LZ77, since it constructs a dictionary which tends 
to grow and fill up during compression. Actually this happens all the time for big inputs 
like in our application scenario, and the common methods to overcome such issue (see 
\cite{salomon2004data}) do not permit to gain the most advantage from the high similarity 
of genomic sequences.

The first self-index based on LZ77 was presented by \cite{kreft2011self}: 
it offers good compression ratio and search performance, but its internal data structures 
were not designed to explicitly handle a collection of data items. This index also does 
not exploit the fundamental requisite of our application domain, that is the compression 
of genomic sequences relative to a reference sequence.

The first attempt to compress a collection of individual genomes with respect to a reference 
sequence was made by \cite{brandon2009data}. That work, like those of \cite{kuruppu2010relative} 
and \cite{kuruppu2011optimized}, aimed to build data structures suitable to efficiently compress 
the collection, while allowing fast random access to parts of it. Pattern search still remained 
an open question.	

The problem of efficiently searching for patterns in a such index was addressed and 
resolved later by \cite{wandelt2013rcsi}, but some of the data structures used therein do not 
allow the encryption of sequences related to different individuals with distinct keys.

\section{System and Methods}
\label{sec:systemandmethods}
In the human genome, over 3 billion DNA base pairs are coded in just 23 chromosome pairs. DNA
has a variety of known roles including protein-coding genes, non-coding RNA genes, regulatory 
sequences, repeat elements, and extensive regions which remain unknown. When genes 
are expressed, information from a given pattern in a chromosome is used to synthesise the gene 
product, which could be a protein or functional RNA. Searching for and identifying these
patterns is therefore an important biological challenge, which is at the root of bioinformatics 
and its applications in P4 medicine, including new trends in drug discovery.

The $ER$-index is designed to be the building block of an encrypted database, that stores 
ge\-no\-mic in\-for\-ma\-tion about a possibly large set of individuals. The database is 
intended to be the data repository for an on-line DNA pattern search service, which allows
only authorized users to search for generic sequences in the DNA of specific individuals or 
groups of them.
Roughly speaking, an \textit{Encrypted Referential database} ($ER$-database for short) is a 
collection of $ER$-indices intended to offer this kind of service; its access is managed 
through portfolios of secret keys related to a population of individuals and a set of database 
users: 

\begin{definition}{\textbf{Encrypted Referential Database}}
Let $R=\{R_j: j\in J \subseteq \{1,\ldots,22,X,Y\}\}$ be a set of reference sequences for human 
chromosomes.
Let $I=\{I_1,\ldots,I_l\}$ denote a set of individuals and $S=\{S_{ij} \mid (i,j) \in I \times J\}$ 
be a set of genomic sequences, where $S_{ij}$ is the sequence of individual $I_i$ related to 
chromosome $R_j$.
An \textbf{Encrypted Referential Database} (\textbf{ER-database}) for $I$ with reference 
$R$ is a tuple $$D=\{I,R,K,U,ER,P\}$$ where:
\begin{itemize}
 \item $K=\{k_1,\ldots,k_l\}$ is a set of randomly-generated, symmetric encryption keys so 
 that $k_i \in K$ is uniquely and secretly associated  to $I_i$ for $i=1,\ldots,l$;
 \item $U=\{U_1,\ldots,U_r\}$ is a set of database users, where each $U_r$ is allowed to 
 access only to the sequences of a subset of the individuals in $I$;
 \item $ER$ is a set of $ER$-indexes for the population $I$, each one relative to a different 
 chromosomic reference sequence in $R$; 
 \item $P$ is a mapping from $U$ to $I$ that, for each user $U_r \in U$, indentifies the 
 individuals in $I$ whose access is granted to the database user $U_r$. 
\end{itemize}
\end{definition}
A simple implementation provides for an $ER$-database hosted by a file system directory, named 
the \textit{database root}. The database root  contains the database catalog \textit{catalog.xml}, 
which lists all the individuals, users and reference sequences composing the database. 
Moreover, in the database root there are the subdirectories \texttt{references} and \texttt{indexes} 
containing the sets $R$ and $ER$, respectively; and the subdirectory \texttt{security}, which contains 
the key portfolios of the database users in $U$.
The \textit{key portfolio} for a database user $U_r \in U$ contains only the symmetric keys
related to the individuals in $I$ whose genomic information $U_r$ has been granted access; it is 
handled with asymmetric encryption techniques, and encrypted with the $U_r$'s public key so that 
only $U_r$ can read its content by using his/her private key.

\begin{figure}
\begin{center}
\includegraphics[width=0.90\textwidth]{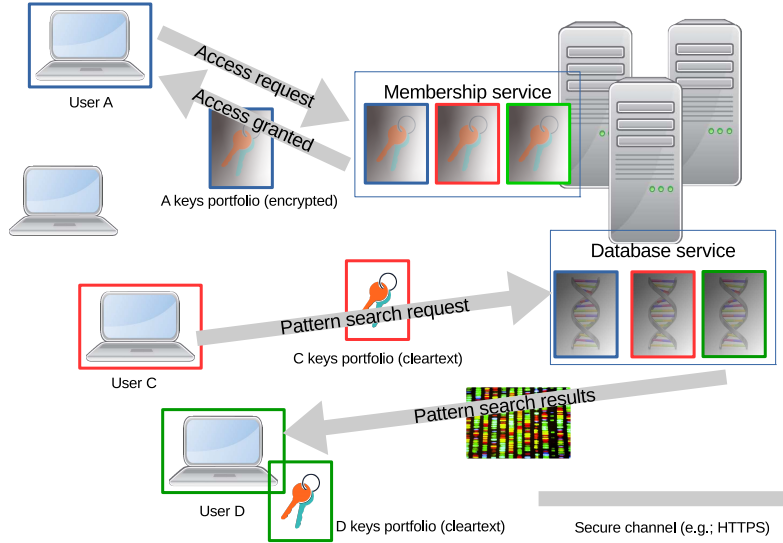} 
\caption{A client-server architecture with a two-tier access control for an ER-database biobank service. 
The database service provides access to genomic information in the form of $ER$-indices, which are stored 
encrypted with multiple user-specific, server-side generated keys. The membership service stores the above
keys as key portfolios of authorized users, encrypted with their public keys. Database users can execute 
pattern searching operation only if they: (i) are successfully authenticated to the membership service, 
and (ii) have the private key for decrypting their personal key portfolios.}\label{fig:er_scenario}
\end{center}
\end{figure}

An ER-database can be deployed as on-line service in different ways, each way corresponding to a given 
architecture with its own threat model and security objectives. In the context of massive data storage or 
processing, a common scenario nowadays is that of a service offered by a third party via a Web interface, 
restfull APIs and, possibly, a dedicated mobile app. This is the case for human genomes too, with currently 
more than a hundred clinical and research oriented services plus over a dozen of consumer genomics service 
providers \cite{genomesource}. 
Some of these providers, especially in the context of consumer genomics, already perform encryption for 
sensible data at rest, although through methods and practices often not well documented for users (as for 
the major provider \textit{23andMe} \cite{23andme}). However, as already discussed in the Introduction, 
encryption is overall rarely adopted by these providers, because of current regulations, the widespread 
perception that it is computationally too expensive and complex, combined with the fact that there is an 
underestimation of the risks related to both the inference of personal information from anonymized data 
and data theft. 

\noindent
In the following of this section we first sketch an architecture which aims at overcoming the above 
limitations, offering a data theft resistant service without appreciable loss of performance in both 
compression and pattern search speed. Then, we discuss the advantages of this architecture compared 
to the current state of the art in relation to a suitable threat model. Finally, we explore alternative
approaches, both in terms of computing models and cryptographic technologies, illustrating their pros and 
cons compared to the proposed solution, and their practicality in the context of our application domain.

\subsection{Service deployment}
\label{sec:service_deployment}
A client-server architecture for offering the $ER$-database service to a set of remote users is sketched 
in Figure 1. One or more computers --  equipped with fault-tolerant RAM and disk storage, and properly 
configured for storage replication, back-up and fault recovery -- host the database root, possibly making 
use of a distributed file system. The service is accessible to authorized users only via protected channels 
(e.g., http over TLS and/or SSH). In a portable, microservice-oriented architecture the database service 
could be deployed on-premise or in a Cloud environment thanks to a set of armored Docker containers 
\cite{merkel2014docker} (e.g., Linux Alpha with a ZFS \cite{bonwick2003zettabyte} file system), orchestrated
via Swarm or Kubernetes \cite{dockerdocs}.

A \textit{membership service}, possibly implemented in a separate environment, allows users to register 
and authenticate to the service, managing their user profiles and authentication credentials. Depending on 
the security requirements, this service can enforce appropriate authentication mechanisms (e.g., 
multi-factor/multi-channel authentication), and make use of specific tools (e.g., privilege access management 
(PAM) systems \cite{PAM}).

Users access the database service by enrolling and authenticating through the membership service, after that 
they can make use of their private keys (locally stored on their devices) in order to process the genomic 
information (i.e., the set of individuals in $I$) they have been granted access thanks to their user profiles. 
After a successfull authentication of the user $U$, the membership service sends to $U$ her key portfolio, which 
is stored and transmitted encrypted with $U$'s public key. Then $U$ can send to the database service the pattern(s) 
she is searching for, alongside with the keys required to decrypt the genomic data where the search has to
be performed. After which the required genomic data is loaded in memory by the server, decrypted using the 
symmetric keys provided by $U$'s key portfolio, and processed as required by $U$ during her working session. 
The results are returned to $U$ via the session connection, and all data are erased from memory when the session 
is terminated. The symmetric keys used by the database service to protect the genomic datasets and their related 
auxiliary structures (i.e., the three $EB+$ trees implemented in the $ER$-index, see Section \ref{sec:encrypted_referential}) 
can be periodically updated by the system in a way completely transparent to users, and according to a scheduling 
policy designed for a good trade-off between security and computational overhead in function of the activity and 
life-cycle of users' profiles. 

\subsection{Threat model and security considerations}
\label{security_considerations}
In the context of the provision of genomics services as detailed at the beginning of this section,
an appropriate threat model is that which assumes that a service provider acts as a \textit{trusted 
third party} both for the set of individuals enrolled in the genomic database and the (possibly 
different) set of database users. The genomic material must be firsty acquired and processed, and
then digitized, stored and queried for, in full compliance with the regulations in force concerning 
the availability, integrity and confidentiality of personal and sensitive data. In this respect, 
both genomic data providers and consumers place their trust in the service provider, turning to 
supervisory authorities if they believe that it has violated the current legislation on the data
protection. Because of the way, tools and high skills required for genomic data acquisition, the 
scenario of a \textit{fully trusted} service provider is the only one conceivable in practice for
data providers (i.e., individuals to whom the genetic material belongs): as a matter of fact, when 
you send your saliva sample in a special test tube to screen your DNA, you are giving the recipient 
your complete trust in the storage and processing of your full genome and personal data.\\
A less stringent model of trust can instead be assumed in relation to data consumers: in this 
context it is appropriate to suppose a \textit{honest-but-curious} service provider: 
processing is carried out correctly based on the queries performed by users, and results 
are returned to the right requester, but the service could acquire or disclose to a third party 
queries and related results because of their commercial value. Although data consumers are supposed 
to pay for the DNA sequences search service, there could be always the possibility that insiders 
(e.g., the database administrator) started a lucrative business by disclosing this information to 
unauthorized parties. Last but non least the possibility of external attacks -- perpetrated by 
users through escalations of privileges, or remotely thanks to the exploitation of bugs or system 
configuration errors -- cannot be excluded.\\
The \textit{honest-but-curious} trust model will be applied also for database users: they could
surely be interested in learning more about genomic data than what their authorization profiles 
consent, but they are incentivized to follow the protocol since they pay for the service. 
Actually, this assumption rules out possible threats like user collusion and sale of access 
credentials to third parties; however, their consideration fall outside the evaluation of our core 
proposal and rather concern the membership service and complementary protection functionalities like 
auditing and behavioral analysis.\\   
Similarly, threats concerning the network will non be discussed here, since they fall outside the 
evaluation of our proposal. We will just assume that inbound and outbound communications with both 
the membership service and the database services are secured through suitable network protocols like
TLS and infrastructures like PKI. In particular, we will assume the proper functioning of the 
membership service, so that: (i) unique and authentic public keys are associated only to authorized 
user; (ii) symmetric key portfolios are correctly generated and updated on the basis of the 
access permissions granted to the database users with respect to individual genomes populating
the database, and; (iii) users' key portfolios are stored encrypted on disk using their
authentic public keys.\\  
Finally we will assume, as it is customary in security analysis which are not specifically 
crypto-oriented, that encrypted data can be accessed only by knowing or guessing the decryption 
key. 
All the above results in a honest-but-curious, computationally bounded adversarial model: an 
adversary is supposed to try to learn information she is not authorized to access, but 
without deviating from system workflows nor comprimising system functioning; moreover, provided 
that keys are properly chosen, encryption is unbreakeable.

\noindent
In the above model, threats concerning data confidentiality can occur because of: 
\begin{enumerate}
\item vulnerabilities affecting the software or the operating system installed on the servers;
\item unauthorized readings of data by internal personnel; 
\item lack of diligence in the management of access credentials (i.e., secret keys, passwords), 
both at the server and client side.
\end{enumerate}
If compared to a cleartext database, the $ER$-database offers much higher protection against 
all the above threats. The architecture previously sketched generates encryption keys
in a random fashion without user intervention, and these keys -- with the sole exception of 
the system key used to encrypt the additional data structures which composes the $ER$-index 
-- are stored encrypted thanks to users' public keys. Thus a successfull attack against data 
stored on disk at the server could just expose the system encryption key, provided that this 
key is not stored in a hardware security module or protected via software (e.g., generated at 
runtime during system startup and stored only in main memory). However, getting the system key 
gives only information concerning metadata like the number of individuals enrolled in the database, 
the block size, and the offsets of LZ77 factorizations; whereas these last are stored encrypted on
disk thanks to user keys.
An attacker who has no access to any user key can only infer which factors belong to which 
individuals; neither the content nor the position of factors with respect to the reference 
sequence or the individual genomes can be desumed. 
Section \ref{sec:encryption} illustrates how encryption is implemented in the $ER$-index,
detailing how system and user (symmetric) keys are actually used.

An $ER$-database offers more protection also if compared to encrypted genomic databases which 
are not self-indices and/or do not have a built-in encryption. Indeed, a pattern search 
within these databases typically require the decryption and decompression of an amount 
of data which does not fit in main memory and has to be temporary stored in cleartext on 
disk. On the contrary, the $ER$-database has been designed and implemented in order to
operate on cleartext solely in RAM, so that an attacker must be able to perform a memory
dump in order to get decrypted information, a task much more challenging than reading
data from disk.

As respect to client-side security, the architecture previously sketched offer more protection 
against \textit{impersonation} if compared to plaintext databases or database which are encrypted 
with server-side managed encryption keys, as in \textit{23andMe}. Indeed, it implements a two-stage, 
two-factor authentication where the private key required to unlock the key portfolio must be used 
after a successful login to the membership service, as detailed in Section \ref{sec:service_deployment}.

\subsection{Current limits and alternative approaches}
Although our proposal offers significative advantages in performance and security with respect to current
implementations and practice of genomic pattern search services, it has the drawback of requiring a certain
amount of trust in the way genomic data are protected server-side after their acquisition, expecially
during their processing in response to users' queries. The main limitation is that data have to be first 
decrypted before being queried for, thus there are chances they are stolen by insiders or outsiders who
had got access to the RAM through memory dumps, covert channels and/or system bugs.
A common and frequently adopted workaround for high-end, secure servers is trusted hardware like Software 
Guard Extensions (SGX) by Intel or Silicon Secured Memory (SSM) by Oracle, which isolates application data
that resides in main memory, thus supporting data confidentiality and secure computation of the results
even if the machine is compromised. 
However, it is relevant to consider whether confidentiality for the data in RAM could be achieved by changing 
the way the $ER$-index is constructed, or how computations on it are performed. 
In this respect, an objection that arises spontaneously, and which is supported by the huge developments 
in modern cryptography in the last twenty years, is because the $ER$-index was not encrypted thanks to one 
of the modern encryption schemes that allow processing directly on the cipher text.
Another relevant question concerns alternative ways to perform processing, so to distribute computing tasks
among different parties and/or devices and avoid a single point of failure in terms of exposition
to data theft. We will articulate our considerations with respect to the three most promising alternative 
approaches, that is Searchable encryption, Homomorphic encryption, and Secure multi--party computation.

\subsubsection{Searchable encryption}
A \textit{searchable encryption} (SE) scheme \cite{bosch2014survey} allows a server to search in encrypted 
data on behalf of a client, without learning any information about plaintext data. This is usually achieved 
thanks to the client-side creation of a searchable encrypted index, where some \textit{keywords} are extracted
from the documents stored in the database and then they are encrypted or hashed so to give rise to a  
searchable ciphertext. The database is encrypted by the client and it is loaded on a server alongside 
with the searchable encrypted index constructed before. After that, the client can generate \textit{trapdors}, 
which are predicates on \textit{keywords} encrypted so that the server can search the index, see  whether the 
encrypted keywords satisfy the predicate, and return the corresponding encrypted documents.\\
Unfortunately, this model of computation is far from being usable in our application scenario. First of all,
the set of clients (i.e., database users) is generally very different from that of data owners; allowing 
clients to access cleartext genomic data of non-anonymised individuals would violate the assumed trust model 
and all current regulations on sensitive data. And, last but not least, SE schemes are only applicable for
a-priori, well defined sets of keywords; on the contrary, our application scenario requires the ability to 
search for \textit{generic} patterns, with the only restriction that they should have some significance in
genomic analyses. As a matter of fact, sequence lengths usually involved in these analyses would turn out
in huge and unmanageable searchable indexes. For example, DNA regulatory sequences are 5 to 20 base pairs 
long, and can be exact or variable \cite{durbin1998biological}; considering all the possible combinations 
would result in more than $10^{12}$ keywords. Things are no better in applications like \textit{electronic 
polymerase chain reaction} (e-PCR), a computational procedure that is used to search DNA sequences for 
sequence tagged sites, each of which is defined by a pair of \textit{primer} sequences and an expected PCR 
product size. Primers are indeed usually designed to be in the order of 18--22 nucleotides in length 
\cite{van2008principles}.  

\subsubsection{Homomorphic encryption}
After the introduction in 2009 of the first concrete scheme supporting arbitrary computations on encrypted 
data \cite{gentry2009fully}, \textit{fully homomorphic encryption} (FHE) has experienced rapid developments. 
Last generation FHE schemes are still computationally consuming, however, and it is believed to need a long 
way for the practical use of FHE schemes on large datasets. This is true in particular for pattern matching, 
where exact or inexact pattern searches in ciphertext can be performed by homomorphically computing the 
\textit{edit distance} between the searched string and every equal-length string in the dataset. Also using \textit{somewhat homomorphic encryption} (SHE) schemes -- which allow a limited number of divisions to be 
carried out, but are much more efficient since they do not employ the \textit{bootstrapping} noise-management technique -- pattern matching in ciphertext is practically infeasible if performed naively ``bit-a-bit'', 
without any ad-hoc encoding technique. 
For example, computing with an SHE scheme at 80 bit security the edit distance between two strings of 8 
nucleotides required 5h 13m on an Intel Xeon i7 at 2.3 GHz with 192 GB of RAM, and the same task would have required more than 45 times that value using an FHE scheme \cite{cheon2015homomorphic}.\\
Notably, \cite{naehrig2011can} introduced cyphertext coding techniques that, when coupled with the SHE given 
in \cite{brakerski2011fully}, highly boost performance in the computation of specific functions.
Later, by extending the \textit{packing technique} given in \cite{naehrig2011can}, \cite{yasuda2013secure} 
found an efficient method to search for all the Hamming distances between two strings under a given threshold.
This method can be used to efficiently solve some inexact pattern matching problems in encrypted genomic data; however, it unfortunately does not fit our application scenario and threat model. It indeed assumes that the 
database users generate the public-private key couples required to homomorphically encrypt and decrypt data, respectively, and that they perform pattern matching on behalf of data owners. Suppose that a database user $U$ has got permission by a data owner $I$ to perform pattern matching on her genomic data $D_I$, and that pattern matching computations are performed by a third-party, honest-but-curious server (e.g., in a Cloud environment).
Then $U$ computes a public-private key couple, giving the public key to $I$ so that she can upload $D_I$ encrypted on the server. When $U$ needs to search for a pattern $P$ in $D_I$, he encrypts $P$ with the same key and uploads the ciphertext on the server, which performs its computations and returns the encrypted result to $U$. 
Finally, $U$ can desume the matching positions by decrypting the result with his private key. Actually, by decrypting the result $U$ will learn all the Hamming distances between $D_I$ and $P$ up to a
threshold $t$, which determines the size of the plaintext space and must be chosen consistently with other 
system parameters in order to get an adequate level of security. Experiments in \cite{yasuda2013secure} show
that $t$ is usually large enough to allow $U$ to infer all the data $D_I$. Thus, the above approach is not compliant with our assumption of honest-but-curious database users; rather, it well fits with the scenario 
where database users are the data custodians that, alongside with data owners, externalize computations in 
the Cloud.

Another severe limitation of homomorphic encryption concerns more generally its practicality in the context of compressive genomics. The more operations (expecially divisions) will be operated on cyphertext, the more will be the size of the resulting ciphertext. The packing technique given in \cite{yasuda2013secure} results in size increase rate in ciphertexts of about 120 times, and rates an order of magniture greater can affect non optimized schemes \cite{yasuda2013practical}. These values greatly downgrade compression performance. 
    
\subsubsection{Secure multi-party computation}
Distributed computing classically deals with questions of computing under the threat of machine crashes
and other inadvertent faults. Instead, in \textit{Secure multiparty computation} it is assumed that a protocol execution may be
threatened  by an external entity, or even by a subset of the participating parties, with the aim of learing private 
information (honest-but-curious adversary model), or also cause the result of the computation to be incorrect (fully 
malicious adversary model).
The definition of security given in both models (see for example \cite{lindell2009secure}) seems to be very restrictive, in that no adversarial success is tolerated; quite surprisingly, however, powerful \textit{feasibility} results have been established since by
the nineties \cite{yao1986generate, micali1987play, goldwasser1988completeness}, demonstrating that in fact, any distributed computing task can be securely computed.
However, secure multiparty protocols are often not usable in practice: on large datasets, requiring a relatively high number of computations like modular exponentiation per bit of the input is completely infeasible. For example, completing a single secure
computation of 4095-bit edit distance required more than 8 hours using a highly parallel implementation based on garbled circuits 
running on 512 cores of a cluster computer \cite{kreuter2012billion}. 

Things go far better in terms of efficiency for \textit{Secret sharing} \cite{shamir1979share}, which is often used as
building block of secure multi-party protocols. 
A secure secret sharing scheme distributes shares so that anyone possessing a number of shares under 
a given threshold has no more information about the secret than someone with no shares. A possible scheme for the $ER$-index 
would be to distribute the computations among $n$ servers, so that each of them perform pattern matching on $n$ ``faked'' 
sequences that must be recombined to get the true ones. For example, a sequence $S$ relative to a given individual could be 
decomposed in $n$ pseudo-random sequences by: (i) encoding $S$ as a binary sequence; (ii) computing $n-1$ pseudo-random binary 
sequences $X_i$ of the same length of $S$, and; (iii) obtaining the last sequence $X_n$ through the bitwise \texttt{XOR} of $S$ 
with the previous $n-1$ sequences $X_i$, $X_n = S \bigoplus_{i=1,\dots,n-1} X_i$. This way, each server can build an $ER$-index
starting from its own faked sequences, and the reconstruction of the true sequences would require the collusion of $n-1$ 
servers. However, pattern matching -- at least as implemented in self-indices -- is not conservative with respect to this kind of 
data splitting. Thus, the searching operation for a pattern $P$ in $S$ would require a \textit{dealer node} which knows both 
$S$ and at least one  $X_i$, so that it can recompute $P$ for the $i$-th server starting from $S \oplus X_i$. Moreover, the
recomputation of $P$ varies in function of the position in $S$ where it is searched; thus a single user query would result in
an interactive protocol among the dealer and the $i$-th server. At the cost of more workload and communication overhead,
this approach mitigates the exposition of cleartext data in main memory. Indeed, assuming that the dealer chooses the $i$-th 
server in a pseudo-random fashion and that it discard the secret $S$ after the computation of $S \oplus X_i$, an attacker has
to exploit both the dealer and one server in the worst case.

\section{Encrypted referential index}
\label{sec:encrypted_referential}
The $ER$-index is an open-source C++ tool designed to handle an encrypted genomic database.
It is a full-text index consisting substantially in two major components:
\begin{itemize}
 \item a set of relative Lempel-Ziv factorizations, one for each sequence of the collection;
 \item a set of three \textit{B+ trees}, as auxiliary data structures to support encryption 
  and search operations.
\end{itemize}
Both the factorizations and the B+ trees are designed to permit efficient pattern searching, 
while allowing users to search only on sequences in the database to which they were granted 
access.

In order to apply encryption with a small overhead in searching and compression 
performance, each factorization is splitted in a series of blocks of factors, so that 
each of them contains the same number of factors.
Each block is then processed independently, so to produce a compact representation 
whose size depends on the compressibility of the information addressed by its factors. 
Finally, the variable size compressed blocks are independently encrypted from each other 
using the \textit{Salsa20} cipher of \cite{bernstein2005salsa20}. 
Salsa20 was one of the ciphers selected as part of the eSTREAM portfolio of stream 
ciphers (see \cite{babbage2008estream}), and has been designed for high performance 
in software implementations on Intel platforms. 
It produces a keystream of $2^{70}$ bytes from 
a 256-bit key and a 64-bit arbitrary \textit{nonce} which is changed for each new 
run with the same key. It subsequently encrypts a sequence of $b$ bytes plaintext 
by XOR-ing it with the first $b$ bytes of the stream, discarding the rest of the 
stream.

A main point in protecting long-term, sensitive information -- as that provided by 
genomic databanks -- is to provide encryption methods which can outstand advanced 
attacks and next generation computing paradigms and platforms.
As of 2019 there are no known attacks on Salsa20, and the 15-round reduced version of 
this cipher was proven 128-bit secure against differential cryptanalysis by \cite{mouha2013towards}. 
Moreover, according to  \cite{grote2019review}, it is resistant against side channel attacks and 
the new emerging quantum computing platforms.

\subsection{Relative Lempel-Ziv factorization}
Let $S$ be a finite string of symbols over a finite alphabet $A$.
Lempel-Ziv methods consist in rules for parsing $S$ into a sequence of factors, so to
replace repetitions in $S$ by using references of their previous occurrences. Factors
contain indeed references to a \textit{dictionary} of substrings in $S$. The difference 
between the LZ77, LZ78 and the relative Lempel-Ziv factorization is that both the LZ77 
and LZ78 build their dictionary ``on the fly'', putting in it substrings encountered in 
$S$ before the current scanning position, whilst the relative Lempel-Ziv factorization 
obtains its compression by comparing the text to an already existing dictionary.   

In the context of our application domain, $S$ is a genomic sequence of an individual 
belonging to a given species for which a reference sequence $R$ is known. As it is 
well known in Genomics $S$ is very similar to $R$, presenting only a few number of 
mutations, deletions and insertions, often in a percentage not greater than 1\%.
Thus using $R$ to construct the dictionary rather than $S$ can allow a better
compression of $S$: indeed, a given portion of $S$ is more similar to the 
corresponding portion of the reference sequence than to a previously seen 
substring of $S$. This is the basic idea of the so-called \textit{Referential Genome 
Compression}, which can be implemented thanks to Relative Lempel-Ziv factorization.

\begin{definition}{\textbf{Relative Lempel-Ziv factorization}}
Let $S$ and $R$ two finite strings over the same finite alphabet $A$. The 
Relative Lempel-Ziv factorization of $S$ with respect to the reference $R$, 
denoted as $LZ(S|R)$, is a sequence of $n$ factors $$ z_0 \cdots z_{n-1}\ .$$
Each factor $z_j$ $(j=0,\ldots,n-1)$ is a triple $\langle p_j,l_j,mc_j \rangle$, 
where:
\begin{itemize}
\item $p_j$ is the position of the longest substring $r_j$ in $R$ matching the
current substring $s_j$ in $S$ with the exception of its last char;
\item $l_j$ is the length of $r_j$;
\item $mc_j$ (a.k.a. \textit{mismatch character}) is the last character in 
$s_j$, so that $s_j = r_j || mc_j$, where $||$ denotes the concatenation between 
two strings.
\end{itemize}
\end{definition}

\subsection{B+ trees}
B Trees and their B+ variant (\cite{BTRees_DonghuiZhang}) are dynamic balanced trees 
whose nodes contain data values and their related search keys. They are often used for 
databases and file system indexing due to the fast search operation they allow to perform. 
The main difference between B and B+ trees is that the former allows every node to contain 
data values, while in the latter these can be found only in leaf nodes, with every 
other node containing only search keys.

\begin{definition}{\textbf{B+ tree}}
Let $N$ be a positive integer. A B+ tree of \textit{order} $N$ is a tree with the 
following properties:
\begin{itemize}
\item All leaf nodes are in the same level, i.e. the tree is \textit{balanced};
\item Every non leaf node contains multiple \textit{search keys}, stored in increasing 
order, which act as separation values for its subtrees: the left sub-tree of a key 
contains values lower than those of the father key, and the right sub-tree contains 
the values greater than those of its father key;
\item Every internal node contains at least $N$ keys and at most $2N$ keys;
\item If the root node is not a leaf node, it has at least one key and at most $2N$ keys.
\end{itemize}
\end{definition}

B+ trees support efficient updates and exact match queries, which find the values 
associated to a given key. They also permit to do efficiently an operation known in 
literature as \textit{range query}, which finds all the values related to keys 
in the interval $[l,r]$. 

Typically B Trees nodes are stored on secondary storage as fixed size disk pages, 
whose size is a multiple of the hosting file system page size.
In order to increase the number of keys and pointers stored in each page, a compression 
scheme can been applied which takes advantage of the fact that keys in a node are very 
close each other. For the B+ trees implemented in the $ER$-index was used the 
\textit{Invariable Coding} method of \cite{krizka2009benchmarking}, which for each node 
stores the first key and the differences between any other key and the first one, 
using the minimum number of bits required to express the difference of the last key
from the first key.

\subsection{The $ER$-index}
Let $\{S_1,\ldots,S_l\}$ be a collection of sequences corresponding to $l$ different 
individuals. Let $R$ and $R_{rev}$ a reference sequence and its reverse, respectively. 
Let $f_i=LZ(S_i|R)$ be the relative LZ-factorization of sequence $S_i$ with respect 
to $R$. Let $BL_{i,1}, \ldots, BL_{i,bn(i)}$ be the sequence of blocks of factors 
gotten from $LZ(S_i|R)$, where $bn(i)$ denotes the obtained number of blocks 
and each block contains exactly $bs$ factors.
Finally, let $S20(plaintext, key, nonce)$ denote the Salsa20 encryption of $plaintext$
with a 256-bit secret $key$ and a 64-bit $nonce$.    

The $ER$-index stores each of the aforementioned blocks encrypted, using a different 
secret key $k_i$ for each individual and the block number as nonce, so that the 
encryption $E(f_i,k_i)$ of factorization $f_i$ is given by:

$$ E(f_i,k_i) = S20(BL_{i,1},k_i,1) \cdots S20(BL_{i,bn(i)},k_i,bn(i)) . $$

In order to speed up search operations, we have designed the \textit{Encrypted B+ 
tree} ($EB+$ tree), an extended variant of the $B+$ tree data structure. An ordinary 
B+ tree, for each key $k_i$, would simply store all the values $v_{i,0},\dots, v_{i,h_i}$
associated to that key. The requirement is instead to index an entire collection of 
genomic sequence factorizations related to many individuals, whose values are stored 
encrypted and must be associated to the right keys.
Thanks to an $EB+$ tree each single factor of the encrypted collection 
$\{E(f_i, k_i)\}$ is associated to the right identifier $i$ and encryption key $k_i$.

The three $EB+$ trees composing the $ER$-index (see Definition \ref{def:erindex}) are
encrypted thanks again to the Salsa 20 cipher but using system-wide secret keys, since 
they are data structures accessed by all users.

Before performing encryption, we use \textit{Invariable Coding} for both node search 
keys and values, but in a different way than in \cite{BTRees_DonghuiZhang}. 
Indeed, the authors of that work applied compression to arrange more values into 
fixed-size node pages, whereas we used it in order to obtain smaller variable 
length nodes, thus minimizing the overall index size.

\begin{definition}{\textbf{Encrypted Referential index}} 
\label{def:erindex}
An Encrypted Referential index ($ER$-index) for a collection of 
sequences $\{S_1,\ldots,S_l\}$ with respect to a reference sequence $R$ and a set 
of encryption keys $\{k_1,\ldots,k_l\}$ is a self-index consisting of:
\begin{itemize}
 \item the encrypted relative Lempel-Ziv factorizations of sequences $\{S_1,\ldots,S_l\}$
  with respect to $R$: $\{E(f_1,k_1),\ldots,E(f_l,k_l)\}$; 
 \item a set of three EB+ trees whose search keys are respectively:
      \begin{enumerate}
      \item $sai\_rev_j$, a suffix array index corresponding to a $R_{rev}$ suffix prefixing the 
      reverse of the $j^{th}$ factor referential part;
      \item $sai_j$, a suffix array index corresponding to a $R$ suffix prefixing the $j^{th}$ 
      factor referential part;
      \item $tp_j$, the position of the $j^{th}$ factor referential part in the reference sequence $R$;
      \end{enumerate}
      and whose values are the couples $\langle i, v \rangle$, where $i$ identifies the sequence 
      $S_i$ and $v$ is the Lempel-Ziv factor of the related genomic sequence, encrypted with key $k_i$.
\end{itemize}
\end{definition}
It can be worthwhile to stress here that the three EB+ above are required to allow an efficient 
implementation for the pattern search functions \textit{LocateInternalOccs} and \textit{LocateExternalOccs},
as detailed in Section \ref{sec:pattern_search}.

\section{Factorization algorithm}
\label{sec:factorization}
The factorization algorithm used to build the $ER$-index  slightly differs from that proposed 
by \cite{kuruppu2011optimized} and \cite{wandelt2013rcsi}, as 
the $ER$-index uses a couple of $FM$-indexes to represent the reference sequence $R$ and its 
reverse $R_{rev}$. 
The $j^{th}$ factor is again a triple $\langle sai\_rev\_start_j,l_j,mc_j \rangle$  
of numbers, but in the $ER$-index they have a different meaning:
\begin{itemize}
 \item $sai\_rev\_start_j$ is the $R_{rev}$ suffix array index from which to start 
      the backward scan of $R_{rev}$ in order to obtain the factor;
 \item $l_j$ is the length of the factor, comprehensive of the mismatch character;
 \item $mc_j$ is the mismatch character.
\end{itemize}
In order to speed-up pattern search, the algorithm retrieves also the three auxiliary 
data $sai\_rev_j$, $sai_j$ and $tp_j$ stored as search keys in the corresponding $EB+$ 
trees composing the $ER$-index (see Definition \ref{def:erindex}).
The algorithm, whose pseudo-code is given by Algorithm \ref{alg:Factorization}, uses four 
data structures related to the reference sequence $R$:
\begin{itemize}
 \item the $FM$-index $FM$ of $R$;
 \item the $FM$-index $FM_{rev}$ of the sequence $R_{rev}$ gotten by reversing $R$;
 \item a correspondence table $R2F$, which maps a suffix of $R_{rev}$ to the $R$ suffix starting 
  from the same character;
 \item the reverse correspondence table $F2R$, which maps a suffix of $R$ to the $R_{rev}$ suffix 
 starting from the same character.
\end{itemize}
Given a sequence $S$, Algorithm \ref{alg:Factorization} scans $S$ from left to right and 
at each step it tries to factorize the suffix $S_i$ by searching the maximum-length 
referential factor starting from $i$. For this purpose it scans the BWT of $R_{rev}$ 
through $FM_{rev}$, starting from $S[i]$ and proceeding backward on $R_{rev}$ until a 
mismatch is found.
This backward search gives as result the $R_{rev}$ suffix array range containing 
the suffixes prefixing the reverse of $S_p$; the algorithm choose the first among 
them, as they are all equivalent for its purposes.

The further processing of the algorithm consists in retrieving the auxiliary information 
related to the previously found factor.
The $getPositionInReference$ and $backwardStep$ functions, alongside with the utility 
functions listed in Table \ref{tab:FMfunctions} exactly match the canonical $FM$-index 
implementation \cite{pizzachili}, so they are not reported as pseudo-codes.

\begin{algorithm}[H]
\caption{Factorization algorithm}\label{alg:Factorization}
\begin{algorithmic}[1]
\Function{Factorize}{$S$,$FM_{rev}$,$FM$,$R2F$,$F2R$}
\State $j \gets 0$  \Comment{Current factor index}
\State $l_{max} \gets 0$;  \Comment{Maximum factor length}
\State $len \gets length(S)$;	
\State $i \gets 0$;
\While{$i < len $}
 \LineComment{Retrieve the next factor}
 \State $nrc \gets S[i]$; \Comment{Curr char, not remapped in the FM index}
 \State $l \gets 1$ \Comment{Curr length of the next factor ref part}
 \If{$i < len - 1$ AND $isInRef(FM_{rev},nrc) $}
  \State $lastNrc \gets nrc$;
  \LineComment{Start a backward search on the rev ref index}
  \State $c \gets remap(FM_{rev},nrc)$; \Comment{Remap curr char}
  \State $sp \gets C(FM_{rev},c)$;
  \State $ep \gets C(FM_{rev},c+1)-1$;
  \State $backStepSuccess \gets true$;
  \LineComment{Backward search stops when the ref part includes the last}
  \LineComment{but one char of S OR the next char is not in the ref sequence}
  \LineComment{OR the last backward step was not successful OR the}
  \LineComment{next char is N and the last is not OR viceversa}
  
\algstore{RLZFactorization}
\end{algorithmic}
\end{algorithm}

\begin{algorithm}[H]
\ContinuedFloat
\caption{Factorization algorithm (continued)}\label{alg:FactorizationContinue}
\begin{algorithmic}[1]
\algrestore{RLZFactorization}
 \WhileNoDo{$i+l < len-1$ AND }
   \StatexIndent[4.8] $isInRef(FM_{rev},nrc \gets S[i+l])$ AND
   \StatexIndent[4.8] $backStepSuccess$ AND
   \StatexIndent[4.8] $(lastNrc \neq N$ AND $nrc \neq N$ OR
   \StatexIndent[4.8] $lastNrc = N$ AND $nrc = N)$ 
  \algorithmicdo
    \State $c \gets remap(FM_{rev},nrc)$;
    \State $trySp \gets  C(FM_{rev},c) +$
    \StatexIndent[4.8] $Occ(FM_{rev}, EOF\_shift(FM_{rev},sp - 1),c)$;
    \State $tryEp \gets  C(FM_{rev},c)+$
    \StatexIndent[4.8] $Occ(FM_{rev}, EOF\_shift(FM_{rev},ep),c)-1$;

    \If{$trySp \leq tryEp$}
     \State $sp \gets trySp$;
     \State $ep \gets tryEp$;
     \State $l \gets l+1$;
     \State $backStepSuccess \gets true$; 
    \Else
     \State $backStepSuccess \gets false$;
    \EndIf
    \State $lastNrc \gets nrc$;	
  \EndWhile	    	
    \State $sai\_rev\_pref \gets sp $;
    \State $mc \gets S[i+l]$
    \LineComment{Find $sai\_rev\_start$, $sai\_pref$ and $tp$, as follows:}
    \LineComment{Find $sai$ of $R$ for $sai\_rev\_pref$ of $R_{rev}$}
    \State $sai=R2F(sai\_rev\_pref)$;
    \LineComment{Do $l-1$ back steps on FM index to find $sai\_pref$}
    \For{$i \gets 1$ To $l-1$}
      \State $sai \gets backwardStep(FM,sai)$;
    \EndFor
     \State $sai\_pref \gets sai$;
    \LineComment{Find position $tp$ of $R$ for $sai\_pref$ using}
    \LineComment{$FM$ index marked rows}
    \State $tp=getPositionInReference(FM,sai\_pref)$
    \LineComment{Do one back step on FM index to find $sai\_rev\_start$}
    \State $sai\_rev\_start \gets backwardStep(FM,sai\_pref)$;
    \LineComment{Store the retrieved factor in the factors array}
    \State $factors[j] \gets \langle sai\_rev\_start,l,mc\rangle$;
  \EndIf
  \State $i \gets i+1 $
\EndWhile

\EndFunction
\end{algorithmic}
\end{algorithm}

\begin{table}
\centering
\resizebox{\textwidth}{!}{%
\begin{tabular}{|l|p{6cm}|}
\hline
Function name & Scope \\\hline\hline
\texttt{C(FMindex index ,character c)} & Return the number of occurrences of characters that are lexically smaller than $c$ within the indexed text \\\hline
\texttt{Occ(FMindex index, character c,int k)} & Return the number of occurrences of $c$ in prefix $L[1 \dots k]$, where $L=BWT(T)$ and $T$ is the indexed text \\\hline
\texttt{isInRef(FMindex index, character c)} & Return true if character $c$ appears in the indexed text of a reference sequence\\
\hline
\texttt{Remap(FMindex index, character c)} & Return the internal remapped $FM$-index code for character $c$ (characters are remapped within an $FM$-index, in order to store them efficiently)\\\hline
\texttt{EOF\_shift(FMindex index, int position)} & Macro for going from the first to the last column of the BWT matrix \\\hline
\texttt{backwardStep(FMindex index, int suffixArrayPrefix)} & Perform a canonical backward step on the FM-Index (backward steps are used, for example, in BWT inversion)\\\hline
\end{tabular}
}
\caption{Functions in Algorithm \ref{alg:Factorization} related to $FM$-index implementation (see for example \cite{pizzachili}).}
\label{tab:FMfunctions} 
\end{table}

\section{Pattern search algorithm}
\label{sec:pattern_search}
$ER$-index supports exact pattern matching through Algorithm \ref{alg:LZPatternSearch}.
Before describing the algorithm details, it is appropriate to make some considerations. 
A pattern search operation on a Lempel-Ziv factorization can retrieve two types of occurrences:
\begin{itemize}
 \item \textit{internal occurrences}, which are completely contained in a factor's referential part;
 \item \textit{external occurrences}, also known in literature as \textit{overlapping occurrences}, 
which have at least a character outside of a factor's referential part. 
\end{itemize}
External occurrences can span two or more factors or end with a factor's mismatch character, and 
a solution to find them on $LZ78$ factorizations is proposed in \cite{Navarro:2002:ITU:646491.694972}. 
The search pattern is splitted in all possible ways and, for each split point, the algorithm searches 
for the right side prefix and the reverse left side prefix in two related \textit{trie} data structures 
(\cite{brass2008advanced}). This results in two sets, the factors ending with the pattern's left 
side and the factors starting with the pattern's right side, and the algorithm eventually joins these 
two sets in order to obtain couples of consecutive factors. This approach can be applied also to relative 
Lempel-Ziv factorizations, but in our case it would require two tries for each individual, which is very 
expensive in term of disk space.
Thus the $LocateExternalOccs$ function of Algorithm \ref{alg:LZPatternSearch} follows a similar approach, 
but it makes use of the following less expensive data structures:
\begin{itemize}
 \item the $FM_{rev}$ and $FM$ indexes of algorithm \ref{alg:Factorization}, in order to search
       for the maximal prefix of the reversed left side in $R_{rev}$ and for the maximal right side prefix 
       in $R$, respectively;
 \item a couple of $EB+$ trees to retrieve the factors ending with the maximal prefix of the reversed left 
       side and those starting with the maximal right side suffix, respectively. 
\end{itemize}

As for internal occurrences, the approach in \cite{Navarro:2002:ITU:646491.694972} is based on the fact that
each $LZ78$ factor is the concatenation of a previous factor with an additional character, which is not true 
for relative Lempel-Ziv factorizations. 

\noindent Therefore function $LocateInternalOccs$ of Algorithm \ref{alg:LZPatternSearch} implements an 
original approach which uses once again the $FM$ index of the reference sequence $R$, together with a third 
$EB+$ tree, named $posTree$, whose search keys are the starting positions of factors referential parts in $R$. 
This last $EB+$ tree allows to retrieve the factors whose referential part starts in a given positions range 
of the reference sequence.
The $LocateInternalOccs$ function also uses the auxiliary information $l_{max}$, defined as the maximum 
length of all factors contained in the $ER$-index, which is determined during the factorization process and 
is stored into the index header.

\noindent Since an internal occurrence of the pattern is completely contained in the reference sequence, 
the first step implemented in $LocateInternalOccs$ could have been to retrieve all the pattern's 
occurrences in the reference sequence. However, we have first to check that an individual sequence factor 
containing the reference sequence occurrence really exists, and then retrieve the occurrence positions in 
the individual sequence previously found. 

These issues have been addressed thanks to the following consideration.
Let us suppose that the suffix array interval $[sp, ep]$ is the result of a pattern search on the
reference sequence. The position $tp$ of each interval's element in the reference sequence can be
retrieved using the related reference index marked rows. Given a factor, let also $l$ be the length 
of its referential part, $tpf$ its starting position in the reference sequence, and $m$ the pattern 
length.
A reference occurrence located in $tp$ is also an individual sequence occurrence if and only if:
\begin{equation}\label{eq:individualseq}
  \begin{cases}
    tpf \leq tp \\
    tpf \geq tp + m -l
  \end{cases}
\end{equation}
The first condition is to make sure that a factor's referential part does not start after the first
character of the reference occurrence, while the second that the factor's referential part does
not end before the end of the reference occurrence.
If both the above $tpf$ range bounds were fixed values, the referential part factors could be retrieved 
by performing a range query on $posTree$. 
The lower bound actually is not a fixed value, since it depends from the length $l$ of the referential part 
of the factor, but we can consider the maximum length of all factors $l_{max} \geq l$.
Because of (\ref{eq:individualseq}), the wrong values returned by a range query based on $tp + m - l_{max} \leq tpf$ 
can indeed be filtered out by keeping only those factors complying to $tpf \geq tp + m - l$.

\begin{algorithm}[H]
\caption{Pattern search algorithm}\label{alg:LZPatternSearch}
\begin{algorithmic}[1]
\item[]
\Function{Locate}{$pat$}
 \State $extoccs \gets \Call{LocateExternalOccs}{pat}$; 
 \State $intoccs \gets \Call{LocateInternalOccs}{pat}$;
 \State $occs = extoccs \bigcup intoccs$;
 \LineComment{Sort each individual occurrence by position}
 \State $\Call{Sort}{occs}$;
 \LineComment{Remove any duplicates}
 \State $\Call{RemoveDuplicates}{occs}$; 
 \LineComment{Find each occurrence text position from its factor identifier}
 \LineComment{$factorId$ and its factor offset $FactorOffset$}
 \State $\Call{FindTextPositions}{occs}$; 
 \State \Return $occs$;
\EndFunction

\algstore{LZPatternSearch0}

\end{algorithmic}
\end{algorithm}

\begin{algorithm}[H]
\ContinuedFloat
\caption{Pattern search algorithm (continued)}\label{alg:LZPatternSearchContinue}
\begin{algorithmic}[1]
\algrestore{LZPatternSearch0}
\item[]
\Function{LocateExternalOccs}{$pat$}
\State $occs=[]$
\State $pl \gets \Call{len}{pat}$;
\For{$sp \gets 0$ to $pl-1$}
 \State $splitPointCharacter \gets pat[sp]$;
 \If{$splitPoint > pl/2$}
  \LineComment{The left side part (lsp) is longer than the right side part (rsp),}
  \LineComment{so the factors expected to end with the lsp are less than those} 
  \LineComment{expected to start with the rsp}

  \State $ls \gets substr(pat,0,splitPoint)$; 
  \State $[lsFacts,lsls] \gets \Call{FindLeftSideFactors}{ls}$;
  \LineComment{Scan $lsFacts$ through the individual identifiers $indId$}
  \For{$indId$ \textbf{in} $lsFacts$}
   \LineComment{Get the factorization $f$ from an associative array $fs$}
   \LineComment{with all the individual factorizations}
   \State $f \gets fs[indId]$;
   \For{$factInd$ \textbf{in} $\Call{GetIndRetrFactors}{lsFacts,indId}$}
    \State $fact \gets f[factInd]$;

    \LineComment{Exclude that the left side crosses the starting point}
    \LineComment{of the current factor,since an occurrence of this type} 
    \LineComment{will be found for a preceding split point}
    \If{$fact.len-1 \ge len(ls)$}
     \If{$fact.letter = splitPointCharacter $}
      \State $occ.factInd \gets factInd$;
      \State $occ.factOff \gets fact.len - 1 -len(ls)$;
      \State $occ.endingFactInd \gets factInd$;
      \State $occ.endingFactOff \gets fact.len -1$;
      \State $lsvl \gets lsls$; \Comment{Left side verified length}
      \State $rsvl \gets 0$; \Comment{Right side verified length}
      \If{$\Call{PatRemPart}{f,pat,sp,lsvl,rsvl,occ}$}
       \State $\Call{AddOccurrence}{occ}$;
      \EndIf  		
     \EndIf
    \EndIf	
   \EndFor
  \EndFor

 \Else
  \LineComment{The right side part (rsp) is not shorter than the left side}
  \LineComment{part (lsp), so the factors expectedto start with the rsp} 
  \LineComment{are less than those expected to end with the lsp}
  \State $rs \gets \Call{substr}{pat,splitPoint+1,pl-splitPoint-1}$;
  \State $[rsFacts,rslp] \gets \Call{FindRightSideFactors}{rs}$;
\algstore{LZPatternSearch1}
\end{algorithmic}
\end{algorithm}
 
\begin{algorithm}[H]
\ContinuedFloat
\caption{Pattern search algorithm (continued)}\label{alg:LZPatternSearchContinue}
\begin{algorithmic}[1]
\algrestore{LZPatternSearch1}
  
  \For{$indId$ \textbf{in} $rsFacts$}
   \LineComment{Get the factorization $f$ from an associative array $fs$} 
   \LineComment{with all the individual factorizations}
   \State $f \gets fs[indId]$;
   \For{$factInd$ \textbf{in} $\Call{GetIndRetrFactors}{rsFacts,indId}$}
    \State $fact \gets f[factInd]$;
     \If{$rslp < fact.len-1$}	
      \State $rsvl \gets rslp$;  \Comment{Right side verified length}
     \Else
      \State $rsvl \gets fact.len-1$;
     \EndIf	    
     \If{$factInd > 0$}
      \State $lsFact \gets f[factInd-1]$;
       \If{$lsFact.letter = splitPointCharacter $}
        \State $occ.factInd \gets factInd-1$;
        \State $occ.factOff \gets lsFact.len - 1$;
	\State $occ.endingFactInd \gets factInd$;
        \State $occ.endingFactOff \gets rsvl-1$;
	\State $lsvl \gets 0;$ \Comment{Left side verified length}	
        \If{$\Call{PatRemPart}{f,pat,sp,lsvl,rsvl,occ}$}
	 \State $\Call{AddOccurrence}{occ}$;
        \EndIf  		
       \EndIf
     \EndIf
    \EndFor
   \EndFor

\item[]		
  \EndIf
\EndFor
\State \Return occs;
\EndFunction

\item[]
\Function{LocateInternalOccs}{$pat$}
 \State $occs=[]$
 \LineComment{An internal occurrence occurs certainly in the ref seq}
 \If{$\Call{searchPatInRefIndex}{FM,pat,sp,ep}$}
  \For{$i \gets sp$ to $ep$}
   \State $m \gets \Call{len}{pat}$;
   \State $tp \gets \Call{getPositionInReference}{FM,i}$;
   \LineComment{\textbf{lmax} is the length of the maximum factor in the index}
   \State $facts \gets $
   \StatexIndent[4.8] $\Call{getFactorsInRange}{posTree,tp+m-lmax,tp}$;	
   \For{\textbf{each distinct} $indId$ \textbf{in} $facts$}
    \LineComment{Retrieve the factorization $f$ from an associative array $fs$} 
    \LineComment{containing all the individual factorizations}
    \State $f \gets fs[indId]$;
    
\algstore{LZPatternSearch2}
\end{algorithmic}
\end{algorithm}

\begin{algorithm}[H]
\ContinuedFloat
\caption{Pattern search algorithm (continued)}\label{alg:LZPatternSearchContinue}
\begin{algorithmic}[1]
\algrestore{LZPatternSearch2}
   \For{$factInd$ \textbf{in} $\Call{GetIndRetrFactors}{facts,indId}$}
     \State $fact \gets f[factInd]$;	
     \State $tpf \gets fact.refPartPositionInReference$;
     \State $l \gets fact.len-1$; \Comment{factor refential part length}
     \If{$tpf \geq tp+m-l$}
      \State $occ.factInd \gets factInd$;
      \State $occ.factOff \gets tp-tpf $;
      \State $occ.endingFactInd \gets factInd$;
      \State $occ.endingFactOff \gets$ 
      \StatexIndent[4.8] $occ.factOff + m - 1$;
      \State $lsvl \gets 0;$ \Comment{Left side verified length}		    		
      \State $\Call{AddOccurrence}{occ}$;		
     \EndIf
    \EndFor
   \EndFor
  \EndFor
 \EndIf	
 \State \Return $occs$;
\EndFunction

\item[]
\Function{PatRemPart}{$f,pat,splitPoint,lsvl,rsvl,occ$}
\LineComment{Check if  the occurrence $occ$ is really a whole pattern occurrence,}
\LineComment{updating it if required. It returns true for successful checks.}
\LineComment{This function tries to extend the verified part of the pattern, both}
\LineComment{on the left and the right side, by comparing the yet not verified}
\LineComment{pattern characters with the factors characters preceding and}
\LineComment{following the verified part. For performance the extension is}
\LineComment{made without extracting the full text of the involved factors,}
\LineComment{but scanning the text one character at a time thanks to the}
\LineComment{reverse reference index.}
\EndFunction

\item[]
\Function{FindLeftSideFactors}{$ls$}
\LineComment{Return $(lslsfact, lsls)$, where $lslsfact$ is a list of factors}
\LineComment{ending with the left side longest suffix, and $lsls$ is the left side}
\LineComment{longest suffix length.}
 \LineComment{Find the longest left side suffix that occurs in the reference string}
 \State $ l \gets \Call{findLeftSideLongestSuffix}{ls}$;		
 \State $ lsls \gets \Call{substr}{ls,ls.len-l,l}$;
 \If{$\Call{searchPatRevInRefIndex}{FM_{rev},lsls,sp,ep}$} 
  \LineComment{Find factors whose suffixArrayPosition is in [sp,ep]}
  \State \Return $[\Call{getFactorsInRange}{reverseTree,sp,ep},l]$;
 \Else
  \State \Return [[],0];
 \EndIf
\EndFunction

\algstore{LZPatternSearch3}
\end{algorithmic}
\end{algorithm}

\begin{algorithm}[H]
\ContinuedFloat
\caption{Pattern search algorithm (continued)}\label{alg:LZPatternSearchContinue}
\begin{algorithmic}[1]
\algrestore{LZPatternSearch3}
\item[]
\Function{FindRightSideFactors}{$rs$}
\LineComment{Return $(rslpfact, rslp)$, where $rslpfact$ is a list of factors}
\LineComment{beginning with the right side longest prefix, and $rslp$ is the}
\LineComment{right side longest prefix length}
 \LineComment{Find the longest right side prefix occurring in the reference string}
 \State $ l \gets \Call{findRightSideLongestPrefix}{ls}$;		
 \State $ rslp \gets \Call{substr}{rs,0,l}$;
 \If{$\Call{searchPatInRefIndex}{FM,rslp,sp,ep}$}
  \LineComment{Find factors whose suffixArrayPosition is in [sp,ep]}
  \State \Return $[\Call{getFactorsInRange}{forwardTree,sp,ep},l]$;
 \Else
  \State \Return [[],0];
 \EndIf	
\EndFunction

\item[]
\Function{FindRightSideLongestPrefix}{$rs$}
 \LineComment{Scan backward the reverse index, starting from the first char}
 \LineComment{of the right side and going on until a mismatch is found.}
 \LineComment{Return the right side longrst prefix $rslp$.}
\EndFunction

\item[]
\Function{FindLeftSideLongestSuffix}{$ls$}
\LineComment{Scan backward the straight index, starting from the last char}
\LineComment{of the left side and going on until a mismatch is found.}
\LineComment{Return the left side longest suffix $lsls$.}
\EndFunction

\item[]
\Function{searchPatInRefIndex}{$FM\_index,pat,sp,ep$}
 \LineComment{Perform a canonical backward search on the given $FM$-index,}
 \LineComment{returning the [sp,ep] suffix array range corresponding to}
 \LineComment{pattern $pat$.}
\EndFunction

\item[]
\Function{searchPatRevInRefIndex}{$FM\_index,pat,sp,ep$}
\LineComment{Perform a backward search on the given $FM$-index, starting from}
\LineComment{the first pattern char, then the second char, and so on.}
\LineComment{Return the [sp,ep] suffix array range corresponding to}
\LineComment{pattern $pat$.}
\EndFunction

\item[]
\Function{GetIndRetrFactors}{$facts,indId$}
\LineComment{Return the indexes of factors belonging to the individual indId,}
\LineComment{selecting them from the collection $facts$, which contains factors} 
\LineComment{belonging to several individuals}
\EndFunction

\algstore{LZPatternSearch4}
\end{algorithmic}
\end{algorithm}

\begin{algorithm}[H]
\ContinuedFloat
\caption{Pattern search algorithm (continued)}\label{alg:LZPatternSearchContinue}
\begin{algorithmic}[1]
\algrestore{LZPatternSearch4}

\item[]
\Function{GetFactorsInRange}{$tree,l,u$}
\LineComment{Return factor Ids in $tree$ corresponding to key values in $[l,u]$}
\State $factorIds \gets []$; \Comment{Retrieved factorIds, grouped by individualId}
\LineComment{Search $tree$ for a leaf containing the lower bound of the search interval}	
\State $startLeafNumber \gets \Call{SearchForLeaf}{tree,l}$;  
\If{$startLeafNumber \not = \textbf{NULL}$}   
\State $curLeafNumber \gets startLeafNumber $;
\State $rangeScanEnd \gets false $;
\State $ i \gets 0 $;  \Comment{index of the current examined key within the current leaf}
\While {$rangeScanEnd = false$}  	    
\If {$i > nk -1 $} \Comment{$nK$ is the number of keys in this leaf}
\LineComment{Node number of the last leaf in this tree}
\If {$curLeafNumber = lastLeafNumber$}  
\State $rangeScanEnd \gets true$;
\Else 
\State $curLeafNumber \gets curLeafNumber+1$;
\LineComment{Load the leaf node from disk and decrypt it}
\State $curLeaf \gets getNode(tree, curLeafNumber)$;
\State $ i \gets 0$;
\EndIf
\EndIf
\If {$ rangeScanEnd = false $} 
\State $k \gets currentLeaf->keys[i]$;
\If { $k>u$ }
\State $ rangeScanEnd = true$;
\ElsIf {$k >= l$}
\LineComment{add to $factorIds$ the retrieved factors identifiers,}
\LineComment{grouped by individualId}   				 		   	
\EndIf
\State $ i = i + 1 $;
\EndIf;  	    	  	      	   

\EndWhile	
\EndIf

\State \Return $factorIds$
\EndFunction

\item[]
\Function{SearchForLeaf}{$tree,value$}
\LineComment{Search $tree$ for a leaf containing the given key $value$ lower bound of}
\LineComment{the search interval, starting from the root node. Returns the found} 
\LineComment{leaf node number if that leaf exists, NULL otherwise}
\EndFunction

\end{algorithmic}
\end{algorithm}

\section{Encryption algorithm}
\label{sec:encryption} 
In the following just the symmetric algorithm used to encrypt LZ77 factorizations and their auxiliary 
data structures will be described. Indeed, cryptographic schema and protocols for encrypting key portfolios 
and implementing secure communications (see Section \ref{sec:systemandmethods}) are already largely in use
today to protect data at rest and Internet traffic, respectively, and they fall outside our core contribution.\\
As detailed there, the encryption algorithm itself is the well know Salsa 20 code introduced in 
\cite{bernstein2005salsa20}, whose security has been throughtly tested over time and which is - alongside with 
its variants - more and more used over the Internet by big players like Google.
Our contribution in this respect consists in how the core algorithm and its related security paramenters 
(keys, nonces) are used and engineered in software, in order to improve both performance and security by 
an optimized management of the encryption and decryption processes. These are implemented so to get an 
efficient domain decomposition over the structures and data composing the index, with respect to its LZ77 
factors and the individuals enrolled in the database. This in turn allows to load and decrypt in main memory 
just the portions of the index required for a given pattern search, other than a full exploitation of 
multithreading in the construction of factorizations.

\begin{algorithm}[H]
	\caption{Encryption algorithm}\label{alg:Encryption}
	\begin{algorithmic}[1]
		\item[]
\LineComment{Save the index on disk, encrypting all its data structures with Salsa20.}
\LineComment{Encryption goes on by switching between different encryption contexts, }
\LineComment{corresponding to different encryption keys and nonces.} 
\Procedure{SaveIndex}{$userPortfolio,indexFileName$}
 \LineComment{Save the index header using the system key and nonce 0}
 \State $sk \gets  \Call{GetSystemKey}{userPortfolio}$;	
 \State $indexHeaderEC \gets  \Call{CreateEC}{sk,0}$;		
 \State $\Call{SaveHeader}{indexHeaderEC}$; 
 \item[]
 \LineComment{Save factorizations related to each individual}
 	 \For{$indId$ \textbf{in} $indexedIndividuals$} 	  	    	
	 	\State $f \gets factorizations[indId]$;
	 	\State $ik \gets  \Call{GetIndividualKey}{userPortfolio,indId}$;
		\State $\Call{SaveFactorization}{f,ik}$ ;
	 \EndFor
 \item[]	 
 \LineComment{Save the EB+ trees with different base nonces}	 
 \State $\Call{SaveTree}{reverseTree,10000000}$;
 \State $\Call{SaveTree}{forwardTree,20000000}$;
 \State $\Call{SaveTree}{posTree,30000000}$;

\EndProcedure		

\algstore{Encryption1}
\end{algorithmic}
\end{algorithm}

\begin{algorithm}[H]
	\ContinuedFloat
	\caption{Encryption algorithm (continued)}\label{alg:EncryptionContinue}
	\begin{algorithmic}[1]
 	\algrestore{Encryption1}

\LineComment{Save a factorization, encrypting it with the key of its owner}
\Procedure{SaveFactorization}{$factorization,individualKey$}
\LineComment{Save the factorization header using 0 as a nonce}
	\State $headerEC \gets  \Call{CreateEC}{ik,0}$;		
	\State $\Call{SaveHeader}{headerEC}$; 
\item[]
\LineComment{Save the blocks of this factorization, using the block number as nonce}
\For{$i \gets 0$ to $N-1$}
	\State $block \gets blocks[i]$;
	\State $blockEC \gets \Call{CreateEC}{ik,i+1}$;		
	\State $\Call{SaveBlock}{block,blockEC}$; 
\EndFor	 

\EndProcedure		
\item[]

\LineComment{Save $tree$ on disk with a given base nonce}
\Procedure{SaveTree}{$tree,baseNonce$}
		
	\LineComment{Save the leaf nodes, in consecutive positions}
	\State $currentLeaf \gets tree.firstLeaf$;
	\While {$currentLeaf \not = NULL$}
		\State $\Call {SaveNode}{currentLeaf,baseNonce}$;
		\State $currentLeaf \gets currentLeaf.nextSibling$;
	\EndWhile
    \item[]
	\LineComment{Save recursively all the remaining nodes (``index nodes'')}	
	\State $ rootNode \gets tree.root $;
	\State $\Call {SaveNode}{rootNode,baseNonce}$;
	\item[]	
	\LineComment{Save $tree$ directory, containing the start positions of the tree nodes in} 
        \LineComment{the $ER$-index bytestream}
	\State $directoryEC \gets  \Call{CreateEC}{sk,baseNonce}$;
	\State $\Call {SaveTreeDirectory}{tree,directoryEC}$; 
\EndProcedure		
\item[]

\Procedure{SaveNode}{$node,baseNonce$}

\LineComment{If $node$ is not a leaf, save recur all nodes in the subtree rooted at $node$ }
	\If {$ node.leaf = false $}
		\For {$child$ in $node.children$}
			\State $\Call{SaveNode}{child,baseNonce}$; 
		\EndFor	
    \EndIf

\algstore{Encryption2}
\end{algorithmic}
\end{algorithm}

\begin{algorithm}[H]
	\ContinuedFloat
	\caption{Encryption algorithm (continued)}\label{alg:EncryptionContinue}
	\begin{algorithmic}[1]
		\algrestore{Encryption2}

	\LineComment{Save the part of the node containing only the search keys, encrypting} 
        \LineComment{it with the system key and $baseNonce + nodeNumber+1$}
	\State $sk \gets  \Call{GetSystemKey}{userPortfolio}$;	
	\State $nodeEC \gets  \Call{CreateEC}{sk,baseNonce + node.number +1 }$;
	\State $\Call {SaveKeys}{tree,nodeEC}$;

    \item[]
   
	\LineComment{If $node$ is a leaf, saves the individual-specific factors identifiers related}
        \LineComment{to each search key. Each factor identifier is encrypted using the}
        \LineComment{encryption key of the specific individual it refers to.}
	\If {$ node.leaf = true $}
		
		\LineComment{Initialise the encryption contexts for all the individuals whose factor}
                \LineComment{identifiers appear in this leaf}
		\State $indECs = []$;  \Comment{an associative array}
		\For {$indId$ in $node.indIds$}
				\State $ik \gets  \Call{GetIndividualKey}{userPortfolio,indId}$;
				\State $indECs[indId] = \Call{CreateEC}{ik,baseNonce + node.number +1 }$;
		\EndFor
		\LineComment{Save the factor identifiers}
		\For {$key$ in $node.keys$}
			\For {$value$ in $key.values$} 
			\LineComment{A $value$ is a couple $\langle indId, factorId \rangle$}					
					\LineComment{Save the factorId with the individual's encryption key}
					\State $indEC \gets  $indECs[value.indId];					
					\State {\Call{SaveFactorId}{value.factorId,indEC}};
							
			\EndFor	
		\EndFor	
	\EndIf

\EndProcedure		
\item[]
	
\Function{CreateEncryptionContext}{$key,nonce$}
	\LineComment{Create an encryption context, initialising Salsa20 internal data}
        \LineComment{structures with the supplied key and nonce. Return a data structure}
	\LineComment{containing the encryption context's internal state}
	
\EndFunction

\item[]

\Function{GetSystemKey}{$portofolio$}
\LineComment{Return the system key stored in the user's portfolio.}

\EndFunction	

\item[]
\Function{GetIndividualKey}{$portofolio, indId$}
\LineComment{Return the key used to encipher genomic data of a given individual}

\EndFunction

\end{algorithmic}
\end{algorithm}

\section{Experimental results}
\label{sec:results}
In order to evaluate the $ER$-Index performance, a small ER-database concerning 50 individuals and 
10 users was implemented as described in Section \ref{sec:systemandmethods}.
Using such database, a comprehensive set of tests was performed on different computing platforms to 
measure the compression ratios (i.e., the ratio of the output data size to the input data size as 
defined by \cite{salomon2004data}), and the times required to build the index and search for patterns.\\ 
The results were also compared with a referential index built thanks to the \textit{Sdsl C++ library} 
by \cite{gog2014optimized}, available at \url{http://github.com/simongog/sdsl}. This library provides 
some succint data structures for implementing self-indexes like the \textit{Compressed Suffix Arrays} 
(CSA) by \cite{grossi2005compressed} and the \textit{wavelet tree $FM$-index} ($WTFM$-index) by 
\cite{grossi2003high}, and we extended this last kind of implementation so to manage collections of 
items and report sequence-relative locations.
We have chosen the $WTFM$-index as reference tool since this index is nearly optimal in space apart 
from lower-order terms, achieving asymptotically the \textit{empirical entropy} of the text with a 
multiplicative constant 1 \cite{grossi2003high}. 

The tests were performed on three computing platforms having different resources as follows, in order 
to evaluate the performance of our tool and assess its effectiveness with respect to the reference tool 
for different CPUs, memory sizes and operating environments:
\begin{itemize}
\item \texttt{ser}, a small-size server with an Intel(R) Xeon(R) CPU E5-2697 v2 at 2.7GHz 24 cores processor 
and 180GB of DDR3 1333 MHz memory, running the \textit{CentOS 7} operating system;
\item \texttt{lap}, a laptop with an AMD A10-9600P at 2.4GHz 6 cores processor and 12GB of DDR4 1866 MHz memory, 
running an \textit{Ubuntu on Windows} application on the \textit{Windows 10} operating system with the 
Microsoft-Windows-Subsystem-Linux turned on; 
\item \texttt{clu}, a node of a computing cluster with 2 Intel Xeon CPU E5-2670 at 2.6GHz 10 cores processor 
and 196GB of DDR3 1600 MHz memory, running the \textit{CentOS 6} operating system. 
\end{itemize}

\subsection{Experimental setup} \label{sec:setup}
 
The individual sequences chosen to assess the prototype performance are those related to human chromosomes 11 and
20, for a population of 50 individuals. Chromosome 11 (135,086,622 base pairs) and 20 (64,444,167 base pairs) were 
chosen as representatives of big and small human chromosomes, respectively.
\noindent
The sequences were of two types:    
\begin{itemize}
\item Diploid consensus sequences obtained from the \textit{1000 Genomes Project} 
(\url{www.internationalgenome.org/home}). These sequences were built by starting from the respective 
BAM files and using the \texttt{Samtools mpileup} command along with the \texttt{BCFtools} (\url{www.htslib.org})
and \texttt{vcfutils} utilities (\url{https://vcftools.github.io/index.html}).
\item Pseudo-random sequences obtained by applying single mutations, insertions and deletions to the corresponding 
chromosome reference sequence in the human genome bank HS37D5, a variant of the GRCh37 human genome assembly used 
by the \textit{1000 Genomes Project}. For this purpose \cite{montecuollo2017e2fm} built a tool  which selects, 
with uniform distribution, mutations, insertions and deletions according to the percentages observed on average 
by \cite{mullaney2010small} among different individuals of the human species.
\end{itemize}
Although artificially generated, the second kind of sequences is more appropriate than consensus sequences to 
evaluate real performances, since they are free from spurious symbols caused by sequencing machines errors or 
inaccuracy.\\
For each one of the two above types we considered full length sequences and 1MB sequences, obtained by selecting 
one million basis of those chromosomes. Thus we performed our tests on a total of eight kinds of genomic 
collection sequences, with consensus collections denoted as $11\_1MB$, $11\_FULL$, $20\_1MB$, $20\_FULL$, and 
their artificially generated counterparts identified by the suffix $\_R$. Some tests include also 5MB sequences, 
obtained by selecting five million basis from chromosomes 11 and 20.

The encryption set-up consisted in the generation of fifty 256-bit symmetric keys through the \texttt{openssl rand}
command, plus ten RSA key couples using the \texttt{openssl genrsa} and \texttt{openssl rsa} commands. 
The key portfolio for each of the database users was generated by choosing a subset from the pool of symmetric keys 
and ciphering it with the user's public key.

\subsection{Construction times}
Tables \ref{tab:times_ser} and \ref{tab:times_lap_clu} show times required to construct the $ER$-index on the 
three considered computing platforms; moreover, the first table reports a comparison with the reference tool on 
\texttt{ser}. Similar results were obtained on the other two platforms, showing that times required to build 
the $ER$-index -- except than for some very short (1MB) sequences -- are significantly lower than those for the 
$WTFM$-index , despite the fact that only the $ER$-index implements data encryption. 

\begin{table}[h!] 
\centering
\resizebox{0.98\textwidth}{!}{
    \begin{tabular}{|r|r|r|r||r|r|r|}
    \hline
     & \textbf{20\_1MB} & \textbf{20\_5MB} & \textbf{20\_FULL} & \textbf{11\_1MB} & \textbf{11\_5MB} & \textbf{11\_FULL} \\
    \hline
    \textbf{ER} &11.38 & 33.74 & 455.7 & 23.18 & 42.79 & 1005 \\
    \hline
    \textbf{WTFM} & 20.08 & 132.1 & 2061 & 19.33 & 154.9 & 5693 \\
     \hline
    \end{tabular}
}
\caption{Times (sec) required to build the $ER$-index (ER) and the $WTFM$-index (WTFM) 
on the \texttt{ser} platform.}
\label{tab:times_ser}
\end{table}
This noticeable perfomance has been obtained through our parallel factorization algorithm, which exploits the 
multi-core, hyper-threading architecture of modern processors (see Section \ref{sec:factorization}).
As it can be easily inferred by a comparison of the obtained values, the speed-up increases with the number 
of cores, so it could be greater on higher-end machines with more processor cores.
Note that the full collections have notable sizes (about 2.97 and 6.4 GiB for the $20\_FULL$ and $11\_FULL$ sets, 
respectively), and this resulted in long computing times (about 0.58 and 1.5 hours) on the \texttt{lap} platform. 
However, we believe these last results are not very indicative since probably due to an improper memory management 
by the virtual machine. 

\begin{table}[h!] 
\centering
\resizebox{0.98\textwidth}{!}{
    \begin{tabular}{|r||r|r|r||r|r|r|}
    \hline
     & \textbf{20\_1MB} & \textbf{20\_1MB\_R} & \textbf{20\_FULL\_R} & \textbf{11\_1MB} & \textbf{11\_1MB\_R} & \textbf{11\_FULL\_R} \\
    \hline
    \textbf{lap} & 50.09 & 47.64 & 208775 & 63.64 & 69.32 & 5409862.50 \\
    \hline
    \textbf{clu} & 9.91 & 9.81 & 256.25 & 19.78 & 16.24 & 528.54 \\
     \hline
    \end{tabular}
}
\caption{Times (sec) required to build the $ER$-index on the \texttt{lap} and \texttt{clu} platforms.}\label{tab:times_lap_clu}
\end{table}

\subsection{Compression ratios}
\begin{figure}[htbp!]
\centering
    \subfigure[Chromosome 20]{
        \includegraphics[width=0.98\textwidth]{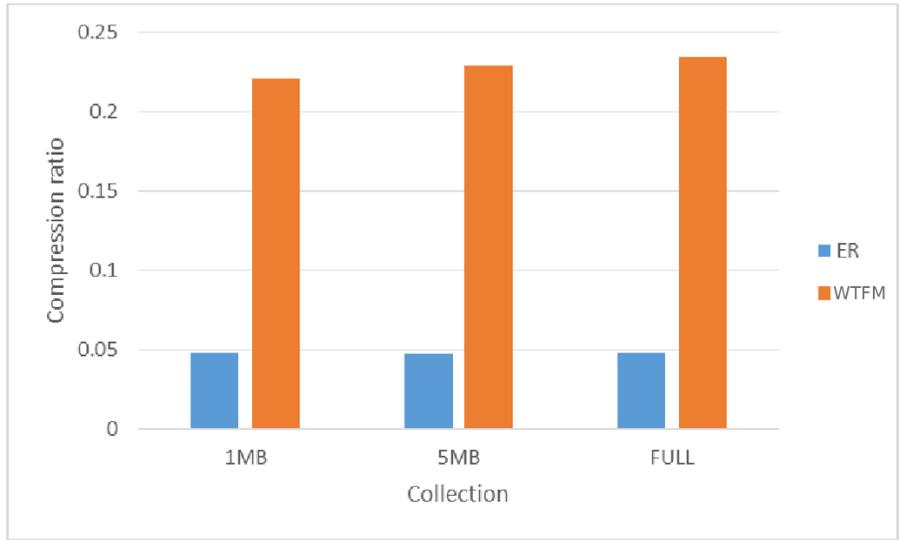}
    }
    \subfigure[Chromosome 11]{
        \includegraphics[width=0.98\textwidth]{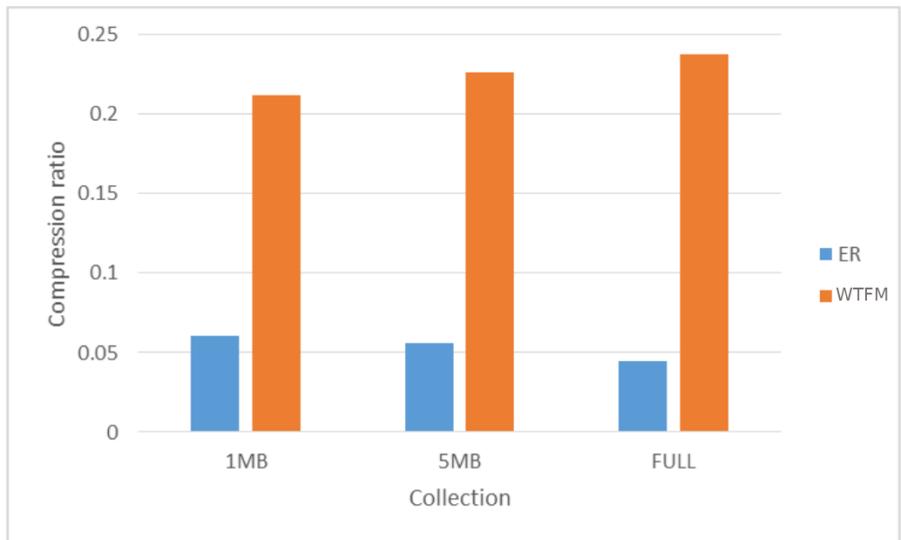}
    }
\caption{$ER$-index (ER) versus $WTFM$-index (WTFM) compression ratios for chromosome 
20 and chromosome 11 collections on the \texttt{ser} platform.}\label{fig:ercr}
\end{figure}

Figure \ref{fig:ercr} reports the compression ratios of the $ER$-index versus the $WTFM$-index on the \texttt{ser} 
computing platform for the collections obtained from the 1000 Genomes Project. 
Obviously, similar results were obtained on the \texttt{lap} and \texttt{clu} platforms, showing that the 
$ER$-index got compression ratios about \textit{four times smaller} than the reference tool.

The compression ratios values on pseudo-random sequences were about half than those for the previous 
sequences: since the sequences created from an algorithm lack of spurious symbols, it is possible 
to find longer matches between the analyzed sequence and the reference one, and achieve a better 
compression performance. Overall this is very good for the $ER$-index, resulting in at least 97\% 
savings in space. For example, the 6.4 GiB of collection $11\_FULL\_R$ resulted in an index smaller 
than 192 MiB. Table \ref{tab:cr_R} reports $ER$-index versus $E2FM$-index and $WTFM$-index compression 
ratios, showing the vastly superior performance of referential compression in the case of large collections,
which in this test were of 50, 75 and 100 items. Table \ref{tab:cr_R} reports also some significant stats
concerning the data structures that compose the $ER$-index. From these stats it can be desumed that the 
memory space required to store a single 60 MiB genomic sequence in the $ER$-index does not depend on the
size of the collection, and it is of about 1800 KiB, where about 1400 KiB are consumed by the auxiliary 
data structure used for encryption and fast pattern matching.    
 
\begin{table}[tbp!] 
\centering
\resizebox{0.90 \textwidth}{!}{
    \begin{tabular}{|r|r||r|r|}
    \hline
    \textbf{Compression ratios} & \textbf{20\_50\_R} & \textbf{20\_75\_R} & \textbf{20\_100\_R} \\
    \hline\hline
    \textbf{$ER$-index} & 0.0288 & 0.0289 & 0.0289 \\
    \hline
    \textbf{$E2FM$-index} & 0.146 & 0.125 & 0.115 \\
    \hline
    \textbf{$WTFM$-index} & 0.237 & 0.215 & 0.196 \\
    \hline\hline
    \textbf{$ER$-index stats (byte)} & \textbf{20\_50\_R} & \textbf{20\_75\_R} & \textbf{20\_100\_R} \\
    \hline\hline
    \textbf{Header} & 437 & 637 & 837 \\\hline
    \textbf{Factorizations} & 20517280 & 30782155 & 41048179 \\\hline
    \textbf{Reverse Tree} & 24285589 & 36524561 & 48630278 \\\hline
    \textbf{Forward Tree} & 24260723 & 36492813 & 48610561 \\\hline
    \textbf{Pos Tree} & 21725297 & 32662668 & 43545681 \\\hline
    \hline
    \end{tabular}
}
\caption{$ER$-index versus $E2FM$-index and $WTFM$-index compression ratios. In this test the collections were 
of 50, 75 and 100 human chromosomes 20, emulated through pseudo-random sequences.}\label{tab:cr_R}
\end{table}
It can be worth to note here that the reported figures are \textit{mean} values obtained by building the index 
more times (we usually performed 18 index builds for each collection, in order to filter out spurious computing time 
values due to umpredictable overheads from other processes running on the same platform). 
As a matter of fact, the multithreading approach causes the operations to be executed in different order during 
the factorization, thus a different order in the creation of the auxiliary data structures. But, since such data 
structures are B+ trees, depending on the order of the insertion of the keys there might be different splits in 
their nodes, resulting in small changes in the size of the index.

\subsection{Pattern search performance}
\begin{figure}[hbpt!] 
\centering
    \subfigure[Mean search times]{       
        \includegraphics[width=0.98\textwidth]{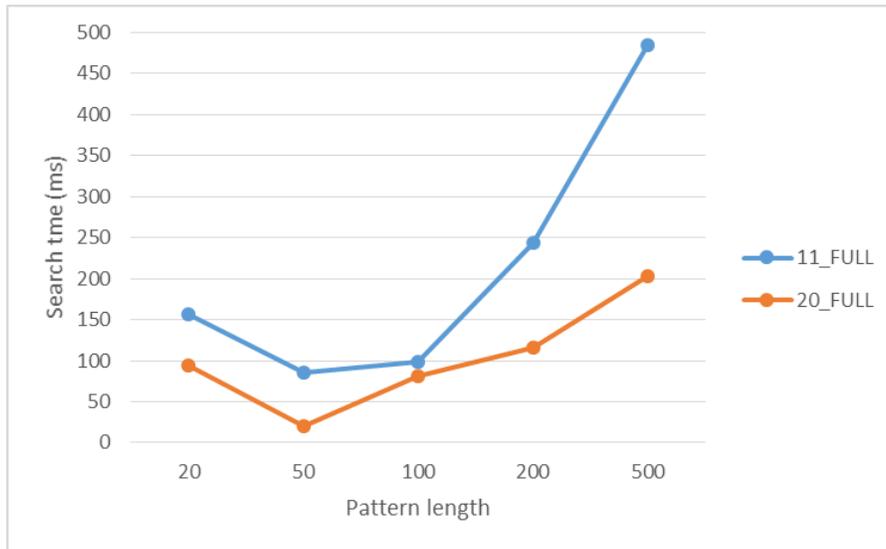}
    }
    \subfigure[Median search times]{
        \includegraphics[width=0.98\textwidth]{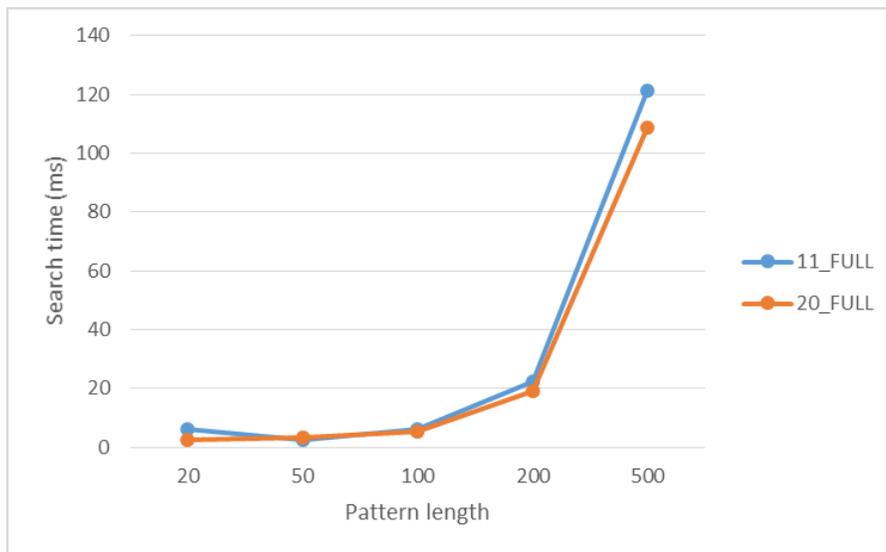}
    }
\caption{$ER$-index mean and median pattern search times (ms) for collections $20\_FULL$ and $11\_FULL$ on the 
\texttt{ser} platform.}\label{fig:st_full_ser}
\end{figure}

For each collection given in Section \ref{sec:setup}, the tests to evaluate pattern search (a.k.a. \textit{locate} 
operation) performance were run as follows:
\begin{itemize}
\item the index ($ER$-index or $WTFM$-index) related to the given collection was selected, and only its header was
loaded in memory;
\item for each pattern length $pl \in \{20,50,100,200,500\}$, $500$ patterns were randomly extracted from the sequences 
composing the collection, and all of them were searched through the index;
\item \textit{mean} and \textit{median} values of the 500 search times and search times \textit{per occurrence} got at 
the previous step were computed;
\item the index was closed, and the next test was performed.
\end{itemize}
Figure \ref{fig:st_full_ser} plots the search time values obtained on the \texttt{ser} platform for collections 
$20\_FULL$ and $11\_FULL$, whereas Table \ref{tab:st_all_lap_clu} sums up the results obtained for the other
collections on the \texttt{lap} and \texttt{clu} platforms.  

\noindent For full collections, pattern search times should be proportional to the number of found occurrences, 
and thus they should decrease with pattern length, since a bigger $pl$  turns out in less chances to find a pattern. 
However, the obtained results clearly show that for such collections search times are higher for $pl=20$ than $pl=50$ but
they increase afterward. This is due to the algorithm used for external occurrences (see Section \ref{sec:pattern_search}), 
since the number of split points checked increases with the pattern length. However, this behaviour could be noticeably 
improved by parallelizing the several split points operations, which are naturally independent from each others. 

Another interesting observation which follows from the obtained results is that median values are significantly smaller than
mean values. This is because of a 10-15\% of outliers, due to some patterns hard to search, or to the fact that some searches
were performed right after the opening of an index, when only a small amount of factorizations and EB+ blocks were loaded in 
memory. Patterns may be hard to search since they have a much greater number of occurrences than the other patterns of the 
same length, or because they span on many short factors so that the left or right side related to some split points are very 
short strings.

\begin{figure}[bpt!]
    \centering
    \subfigure[Mean search times]{
        \includegraphics[width=0.98\textwidth]{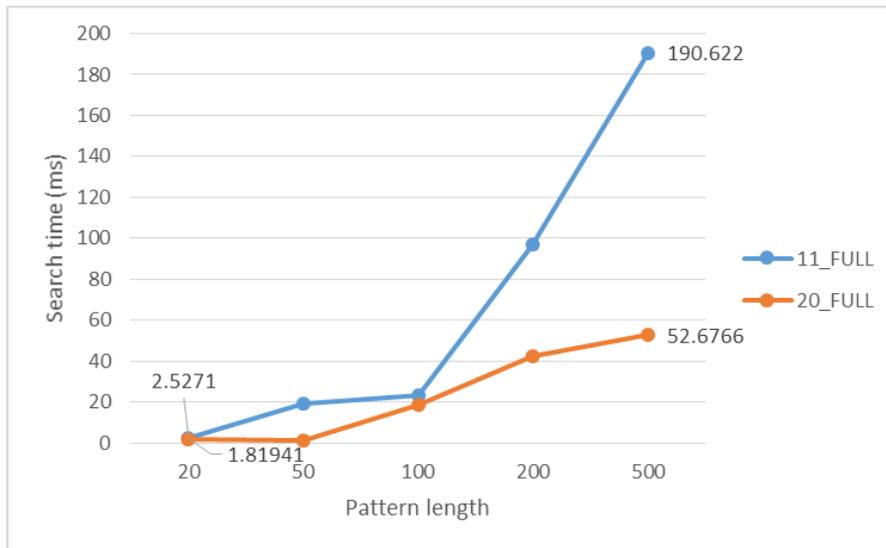}
    }
    \subfigure[Median search times]{
        \includegraphics[width=0.98\textwidth]{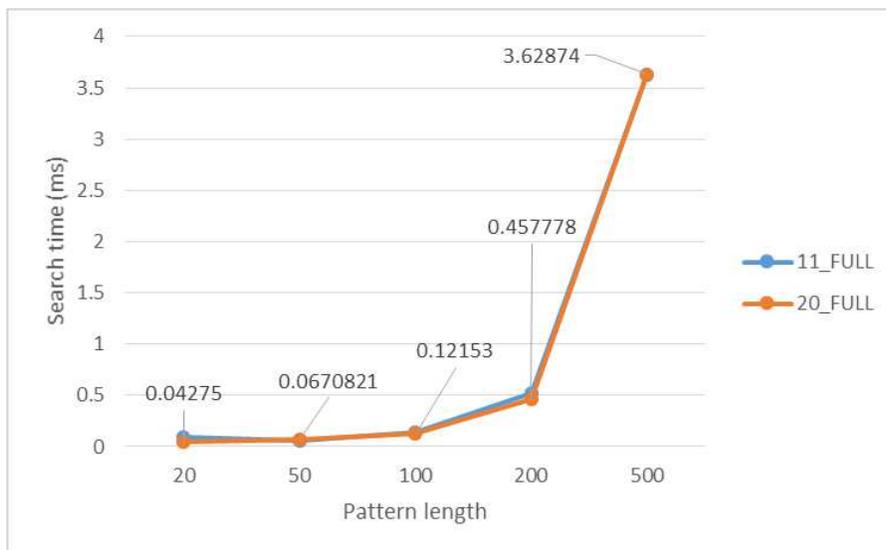}
    }
\caption{$ER$-index mean and median search times (ms) per occurrence for collections $20\_FULL$ 
and $11\_FULL$ on the \texttt{ser} platform.}\label{fig:stpo_full_ser}
\end{figure}

Overall these results show that locate operations are executed very fast thanks to the $ER$-index: also on a small 
computing platform running a virtual machine like \texttt{lap} they take less than half a second in the worst case 
(i.e., looking for 500-basis patterns in the $11\_FULL\_R$ collection of 6.4 GiB). A comparison with the WT$FM$-index 
has shown that this last performs better on patterns with $pl \geq 100$, but slightly worse on short patterns. 
These differences are however of the order of hundredths of a second, so they are ininfluential from a practical point 
of view, except in application scenarios where a massive amount of pattern searches is required.
Since pattern search times are proportional to the number of found occurrences, it is appropriate to look at the mean 
and median values of the search time per occurrence, reported in Figure \ref{fig:stpo_full_ser} for the \texttt{ser} 
platform and the two collections $20\_FULL$ and $11\_FULL$. These results show that mean search times per occurrence 
grow with pattern length, starting from a few milliseconds for $pl=20$ to a maximum of 190.622 ms for $pl=500$ on the 
$11\_FULL$ collection. Note that the curves of the median values related to the two collections perfectly overlap. 
This attests the scalability of the $ER$-index: collections of increasing size can be managed without a significant 
loss in performance.

\begin{table}[htbp!] 
\centering
\resizebox{0.98 \textwidth}{!}{
    \begin{tabular}{|l|r|r|r||r|r|r|}
    \hline
    $pl$ & \textbf{20\_1MB} & \textbf{20\_1MB\_R} &  \textbf{20\_FULL\_R} & \textbf{11\_1MB} &
    \textbf{11\_1MB\_R} & \textbf{11\_FULL\_R} \\
    \hline
    20  & 2.11  & 1.41  &  45.19  & 4.96  & 0.99  & 122.41   \\
    \hline
    50  & 2.71  & 1.89  &  7.89  & 3.07  & 1.73  & 45.72  \\
    \hline
    100 & 7.63  & 6.51  &  12.38  & 8.00  & 5.77  & 19.39  \\
    \hline
    200 & 47.93 & 34.93 &  52.20 & 37.60 & 33.05 & 61.95  \\
    \hline
    500 & 363.47  & 359.59  &  382.79  & 371.36  & 367.98  & 408.45 \\
    \hline\hline
    20  & 0.40 & 0.29 &  4.19 & 0.58 & 0.31 & 8.84  \\
    \hline
    50  & 0.92 & 0.80 & 1.61 & 1.00 & 0.81 & 1.72 \\
    \hline
    100 & 3.04 & 2.78 & 4.25 & 3.18 & 2.87 & 4.70 \\
    \hline
    200 & 13.17 & 12.39 & 14.88 & 13.62 & 13.04 & 16.33 \\
    \hline
    500 & 86.97 & 83.25 & 85.20 & 87.72 & 86.45 & 91.89\\
    \hline
    \end{tabular}
}
\caption{$ER$-index search time mean values (ms) on platforms \texttt{lap} and \texttt{clu}.}\label{tab:st_all_lap_clu}
\end{table}

\section{Conclusion and future work}
\label{sec:conclusion}
Encrypted and compressed full-text indexes are very promising tools for tackling the challenges in terms of
storage requirement, computational cost and privacy posed by human genomic databanks. They indeed allow just to 
decrypt the blocks of compressed data required in a given operation, and to search for patterns directly on
such compressed data, which can results in big time savings if compared to the standard approach 
``decrypt-decompress-search'' achievable using a compressor together with a cipher.
 
In this work we have introduced $ER$-index, the first referential self-index designed to be the core of secure 
genomic databases: it exploits inter-sequence redundancy to get very good compression ratios, and stores the 
sequences of different individuals so that they are encrypted on disk with different encryption keys within the 
same index. 

Our tests have shown that on collections of fifty chromosomes, the $ER$-index achieves compression ratios four 
times better and pattern search times very close to those of the \textit{wawelet tree $FM$-index}, which is a 
self-index nearly optimal in time and space but not implementing referential compression.
For example, search times per occurrency were less than two tenths of a second on a collection of 6.4 GiB, that 
the index compressed and encrypted in less than 192 MiB. 

A multi-threading search strategy and an algorithm for inexact search operations are under investigation and will 
be implemented in a next release of the $ER$-index.

\bibliographystyle{natbib}
\bibliography{AcceptedManuscript}

\end{document}